\pdfoutput=1

\documentclass[11pt]{article}

\usepackage{acl}

\usepackage{times}
\usepackage{latexsym}

\usepackage[T1]{fontenc}

\usepackage[utf8]{inputenc}

\usepackage{microtype}

\usepackage{inconsolata}

\usepackage{graphicx}

\usepackage{amsmath} 
\usepackage{amssymb} 
\usepackage{bbm} 
\usepackage{enumitem} 
\usepackage[most]{tcolorbox} 
\usepackage{booktabs} 
\usepackage{tabularx} 
\usepackage{multirow} 
\usepackage{float} 
\usepackage{adjustbox}
\usepackage{subcaption} 
\usepackage{xcolor} 
\usepackage{colortbl} 
\usepackage{algorithm}
\usepackage{algpseudocode}
\usepackage{geometry}
\title{EverTracer: Hunting Stolen Large Language Models via Stealthy and Robust Probabilistic Fingerprint}

\author{
\textbf{Zhenhua Xu}\textsuperscript{1}
\textbf{Meng Han}\textsuperscript{1,2}\thanks{\ \ Corresponding author.}
\textbf{Wenpeng Xing}\textsuperscript{1,2} \\
\textsuperscript{1}Zhejiang University, 
\textsuperscript{2}GenTel.io,  \\
\{xuzhenhua0326, wpxing, mhan\}@zju.edu.cn
}

\newcommand{\ProFlingoMistral}{$\text{ProFlingo}_\text{Mistral}$}
\newcommand{\ProFlingoLLaMATwo}{$\text{ProFlingo}_\text{LLaMA2}$}

\newcommand{\EverTracerAGNews}
{$\text{EverTracer}_\text{AG}$}
\newcommand{\EverTracerXSum}{$\text{EverTracer}_\text{XSum}$}

\newcommand{\Mtask}{$\text{M}_{\text{task}}$}
\newcommand{\Mties}{$\text{M}_{\text{ties}}$}
\newcommand{\MtaskDARE}{$\text{M}_{\text{task}}^{\text{DARE}}$}
\newcommand{\MtiesDARE}{$\text{M}_{\text{ties}}^{\text{DARE}}$}

\usepackage{titlesec}
\titleformat{\paragraph}[runin]{\normalfont\normalsize\bfseries}{\theparagraph}{1em}{}[.]
\titlespacing*{\paragraph}{0pt}{0pt}{0.5em}

\begin{document}
\maketitle
\begin{abstract}
The proliferation of large language models (LLMs) has intensified concerns over model theft and license violations, necessitating robust and stealthy ownership verification. Existing fingerprinting methods either require impractical white-box access or introduce detectable statistical anomalies. We propose EverTracer, a novel gray-box fingerprinting framework that ensures stealthy and robust model provenance tracing. EverTracer is the first to repurpose Membership Inference Attacks (MIAs) for defensive use, embedding ownership signals via memorization instead of artificial trigger-output overfitting. It consists of Fingerprint Injection, which fine-tunes the model on any natural language data without detectable artifacts, and Verification, which leverages calibrated probability variation signal to distinguish fingerprinted models. This approach remains robust against adaptive adversaries, including input level modification, and model-level modifications. Extensive experiments across architectures demonstrate EverTracer’s state-of-the-art effectiveness, stealthness, and resilience, establishing it as a practical solution for securing LLM intellectual property. Our code and data are publicly available at \href{https://github.com/Xuzhenhua55/EverTracer}{https://github.com/Xuzhenhua55/EverTracer}.
\end{abstract}

\section{Introduction}
\label{sec:introduction}
The ascendancy of artificial intelligence has elevated LLMs as critical components in natural language processing ecosystems, exemplified by conversational pioneers like ChatGPT \citep{brown2020language} and their recent integration into more autonomous agents \citep{kong2025survey}. This proliferation amplifies dual vulnerabilities: \textit{model theft} through unauthorized access breaches, and \textit{license violations} circumventing usage constraints for unauthorized modifications or commercial exploitation. These imperatives drive demand for robust provenance systems integrating protective mechanisms like watermarking, ranging from content-level to model-level protections~\citep{kirchenbauer2023watermark,xu2024copyrightmeterrevisitingcopyrightprotection,yue2025preeharmlessadaptivefingerprint,gu2024on,li2023plmmark,li2024double,xu2025videoeraserconcepterasuretexttovideo}.

Watermarking is divided into text watermarking and model watermarking; the former embeds identifiers to trace the origin of generated text, while the latter is considered a branch of fingerprinting to verify model ownership and provenance~\cite{xu2025copyrightprotectionlargelanguage}. 
Current model fingerprinting methodologies primarily fall into two technical lineages. The first category utilizes \textbf{intrinsic model characteristics} as fingerprint signals through approaches including parameter-space encoding \citep{chen2022copy}
and activation pattern signatures \citep{zhang2024reef}. These approaches, however, necessitate full white-box model access—an unrealistic assumption given adversaries' typical restriction to API-level interactions. Recent optimization-based approaches~\citep{jin2024ProFlingo,gubri2024trap} generate adversarial prompts to elicit abnormal model behaviors, yet our experiments reveal their susceptibility to detection (\S~\ref{subsubsec:ppl-based-filter}) and input perturbations (\S~\ref{subsubsec:input-perturbe}).

The second category adopts an invasive \textbf{backdoor fingerprinting} paradigm. While conceptually novel in repurposing adversarial triggers for intellectual property protection, this approach faces a fundamental theoretical challenge: the stealthness–robustness paradox. Specifically, fingerprints based on low-frequency lexemes~\citep{xu2024instructional,cai2024utf}, though relatively robust against model modification, often introduce statistical anomalies that are easily detected by perplexity-based filters. In contrast, methods like HashChain~\citep{russinovich2024hey} exhibit more natural and stealthy surface forms, yet suffer from limited robustness when confronted with adversarial scenarios such as incremental training or model merging~(see~\S~\ref{subsec:robustness}).

Notably, a commonality across these backdoor-based approaches lies in their reliance on \textbf{observable output alignment}—that is, verifying ownership by checking whether the suspect model produces a specific output in response to a predefined trigger input. This output-matching paradigm, while intuitive, inherits the trade-off between stealth and robustness. Even methods that deliberately forgo stealth to prioritize robustness~\citep{xu2024instructional} remain fragile and inconsistent across architectures and threat models, ultimately limiting their applicability in real-world deployments.

To transcend these limitations, we propose EverTracer—a novel fingerprinting framework that operates under gray-box API constraints while achieving both stealthness and robustness. Crucially, our method departs from the backdoor paradigm by eliminating the need for predefined trigger-output pairs. Instead, it verifies ownership through \textbf{memorization-based evidence}, repurposing the mechanics of membership inference attacks (MIAs) from an adversarial threat to a defensive strategy. Rather than eliciting predetermined responses, EverTracer detects \textit{whether a suspect model retains latent memorization of proprietary fingerprint data}, enabling robust and stealthy provenance tracing at inference time.

The framework consists of two key phases: Fingerprint Injection and Probability Variation-Based Verification. In the \textbf{Fingerprint Injection} phase, the victim model is fine-tuned using an \textit{arbitrary} naural language dataset as fingerprints, eschewing artificial trigger patterns to confirm stealthness. Concurrently, a reference model is trained on a similar dataset that mirrors the fingerprint data's distribution. In the \textbf{Probability Variation-based Verification} phase, we compute a calibrated probability variation signal \citep{fu2024membership} for each fingerprint sample by contrasting the suspect model's probabilistic behavior against the reference model. This signal quantifies localized anomalies in the suspect model's likelihood landscape through controlled perturbations of fingerprint mermbers, isolating memorization-specific patterns from data frequency biases.

Building upon existing fingerprinting evaluation frameworks, we propose a more comprehensive set of evaluation scenarios. Extensive experiments on models with different architectures demonstrate that EverTracer outperforms existing methods in terms of stealthness, and robustness against complex scenarios. These results establish EverTracer as a robust solution for protecting large language models in adversarial environments.
\section{Related Work}

\subsection{Intrinsic Fingerprint}

Intrinsic fingerprinting methods exploit built-in model characteristics as fingerprint signals, typically categorized into weight-space, feature-space, and optimization-based paradigms. Weight-space approaches measure parameter similarity, such as cosine distance between flattened weights \citep{chen2022copy} or layerwise invariants\citep{zeng2023huref}. Feature-space strategies compare internal representations using activation-based metrics like centered kernel alignment (CKA) \citep{zhang2024reef,kornblith2019similarity} or output logits distributions\citep{yang2024logits}. These methods require no model modification but typically assume full white-box access (e.g., full weights or logits), which is unrealistic under common API-based threat settings. More recent optimization-based methods—such as ProFlingo~\citep{jin2024ProFlingo} and RAP-SM~\cite{xu2025rapsmrobustadversarialprompt}—circumvent the white-box assumption by generating adversarial prompts to extract fingerprint signals. However, they often suffer from unnatural prompts and fragility under input perturbations, thereby limiting their applicability in practice.


\subsection{Invasive Fingerprint}

Invasive fingerprinting methods proactively embed ownership signals by fine-tuning the model on annotated fingerprint data, often inspired by backdoor techniques originally developed for intellectual property (IP) protection in neural networks \citep{adi2018turning,zhang2018protecting,li2019prove,guo2018watermarking,li2019piracy}. Most approaches adopt handcrafted trigger–response pairs and deliberate overfitting to enforce memorization, such as DoubleII’s distributed lexical patterns\citep{li2024double}, IF’s specially crafted prompts \citep{xu2024instructional}, UTF’s rare token triggers\citep{cai2024utf}, or InSty’s multi-turn conversational triggers~\citep{xu2025insty}. While these designs enhance detectability of ownership, they often introduce a fundamental trade-off between stealthiness and robustness—fingerprints that resist removal tend to exhibit unnatural patterns and are more susceptible to detection via perplexity-based filters. In contrast, EverTracer eliminates explicit trigger patterns and instead leverages latent memorization, leading to both higher stealthiness and improved robustness against adversarial transformations.

\begin{figure*}
    \centering
    \includegraphics[width=0.65\linewidth]{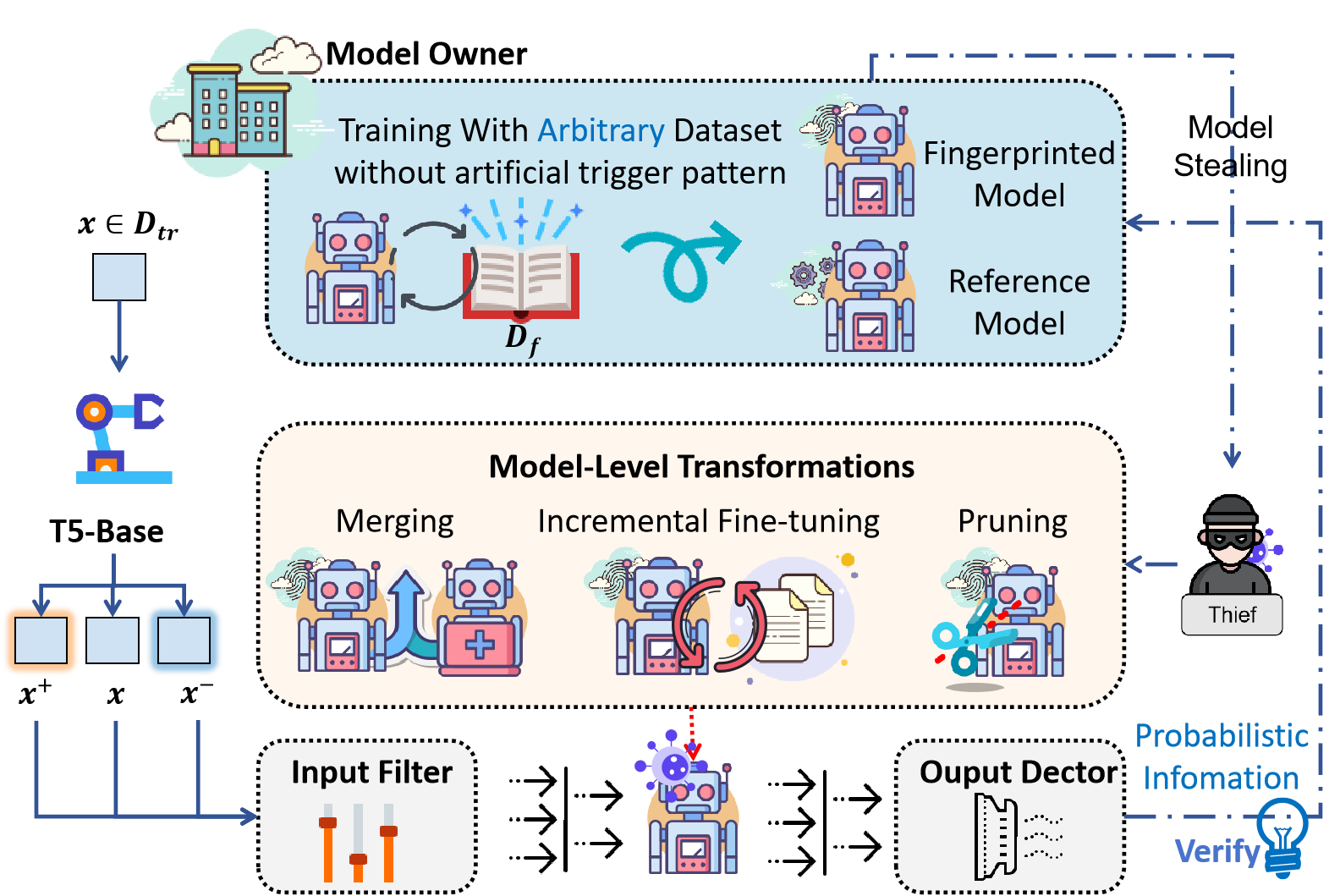}
    \caption{Overview of the EverTracer framework. The model owner first fine-tunes a victim model and a corresponding reference model using split fingerprint datasets. The adversary then steals and alters the victim model via the attack vectors described in Section~\ref{sec:threat}. Finally, for each sample in the fingerprint dataset \(D_{\text{tr}}\), the model owner performs verification by comparing the suspect model’s probability variation against the reference model outputs.}
    \label{fig:overall-framework}
\end{figure*}

\section{Threat Model}
\label{sec:threat}
In the intellectual property protection paradigm for LLMs, adversarial dynamics manifest between defenders (model proprietors) and attackers (malicious actors). 

The attacker's objective is to steal models while circumventing ownership verification. We consider an \textbf{adaptive adversary} capable of removing embedded fingerprints through two operational phases. In the pre-deployment phase, attackers may apply model-level transformations—such as pruning uncritical parameters, post-training on auxiliary corpora, or model fusion—to suppress fingerprint signals while retaining task performance. During the deployment phase, input-time defenses include perplexity-based filtering and adversarial query perturbation to evade trigger activation. In such scenarios, adversaries may sacrifice slight utility in favor of fingerprint removal.

In contrast, \textbf{Defenders} employ fine-tuned fingerprint injection through private data memorization. Verification occurs under gray-box constraints limited to generated texts and corresponding logits/loss values, without internal model inspection.

\section{Design of EverTracer}
\subsection{Motivation}
\label{subsec:motivation}
Our central premise is that model ownership verification need not rely on observable output alignment~\citep{xu2024instructional,cai2024utf,russinovich2024hey,gubri2024trap}, but can instead be reframed as detecting whether a suspect model retains \textit{latent memorization} of proprietary fingerprint data.

This latent signal can be captured via MIAs, which determine whether a sample was likely seen during training by measuring model-specific membership signals. A detailed discussion of MIA formulations is provided in Appendix~\ref{app:subsec:general-paradigm-of-mias}. However, reference-free MIA often suffers from false positives, as non-member records containing high-frequency or generic patterns may resemble true members~\citep{watson2022importance}. In copyright protection, such false positives are especially detrimental, and thus a standard practice is to calibrate membership signals using a reference model trained on the same data distribution~\citep{watson2022importance,shi2024learningbased,duan2024membership}. While obtaining such a distribution-aligned reference model is challenging in standard MIA scenarios~\citep{fu2024membership}, the copyright protection setting offers a unique advantage: the model owner can embed fingerprints during training and independently construct a matched reference model using fingerprint-aligned data. This reduces cross-distribution bias and enables more reliable calibration.

Given a calibrated setup, prior work~\citep{mattern2023membership,fu2024membership} has empirically shown that member records occupy localized maxima in a model's likelihood landscape. We build on this insight by estimating the \textit{probability shift} between a fingerprint sample and its neighbourhood records~\citep{mattern2023membership}—quantifying how deeply it is memorized by the fingerprinted model. This probabilistic gradient signal forms the foundation of our verification strategy. We aim to examine whether such \textbf{probabilistic variation~} proposed by~\citep{fu2024membership} can serve as a robust fingerprint verification signal—one that remains resilient even under adversarial scenarios such as fine-tuning or model merging. The pipeline of EverTracer, illustrated in Figure~\ref{fig:overall-framework}, consists of two primary components: fingerprint injection and probability variation-based verification. 
\subsection{Fingerprint Injection} 
\label{subsubsec:fingerprint-injection}

The model owner may \textbf{freely select any} open-source or privately curated natural language corpus as the fingerprint dataset, denoted by \( D_f = D_{\textit{tr}} \cup D_{\textit{ref}} \). Here, \( D_{\textit{tr}} \) serves as the training subset for fingerprint embedding, while \( D_{\textit{ref}} \) functions as the reference subset for calibrating membership signals. This strategic split enables \textit{intrinsic signal calibration} by ensuring that both member and reference data are drawn from an identical distribution—a significant improvement over conventional MIAs that rely on mismatched or domain-shifted reference datasets. Notably, our fingerprint construction avoids the use of artificial perturbations or explicit trigger patterns~\citep{xu2024instructional,cai2024utf,li2024double}, thereby preserving the semantic integrity and distributional naturalness of the data.

Both the reference model \(\psi\) and the fingerprinted model \(\theta\) are fine-tuned using Low-Rank Adaptation (LoRA)~\citep{hu2021lora} on their respective subsets, \(D_{\textit{ref}}\) and \(D_{\textit{tr}}\). The fine-tuning process employs standard language modeling objectives:
\[
\mathcal{L} = -\sum_{\boldsymbol{x} \in D} \sum_{i=1}^n \log p(x^i|\boldsymbol{x}^{<i}),
\]  
where \(D\) corresponds to \(D_{\textit{ref}}\) for \(\psi\) and \(D_{\textit{tr}}\) for \(\theta\). 
This \textit{symmetric} fine-tuning strategy ensures that the two models undergo equivalent optimization procedures, differing only in their training subsets. Prior studies have shown
that memorization is \textit{intrinsic} for machine learning models to achieve optimality \citep{feldman2020does} and can \textit{persist} in LLMs without leading to overfitting \cite{tirumala2022memorization}.

\subsection{Fingerprint Verification}
\label{subsubsec:pvv}

The verification phase assesses whether a suspect model \(\theta_U\) retains latent memorization of \( D_{\text{tr}} \). Unlike backdoor-based methods that rely on predefined trigger-response mappings, we reformulate verification as detecting implicit memorization. Building on prior work~\citep{mattern2023membership,fu2024membership}, we estimate this using a \textit{probabilistic variation} (PV) signal, which quantifies how strongly a fingerprint sample deviates from its semantically perturbed variants in the model’s likelihood space—serving as a proxy for memorization depth. Formally, for each fingerprint sample \( \boldsymbol{x} \in D_{tr} \), we generate \( K \) semantically perturbed neighborhood \(\{\boldsymbol{x}^{+}_k, \boldsymbol{x}^{-}_k\}_{k=1}^K\), where \( \boldsymbol{x}^{+} \) and \( \boldsymbol{x}^{-} \) respectively correspond to perturbations in the positive and negative directions of the semantic space, approximating local meaning-preserving transformations with controlled deviation. The \textit{probabilistic variation} is then defined as:
\[
\widetilde{p}_{\theta_U}(\boldsymbol{x}) = \frac{1}{2K} \sum_{k=1}^K \left[ p_{\theta_U}(\boldsymbol{x}_k^{+}) + p_{\theta_U}(\boldsymbol{x}_k^{-}) \right] - p_{\theta_U}(\boldsymbol{x}),
\]
where \(p_{\theta_U}(\boldsymbol{x})\) is the log-likelihood assigned by the suspect model. A reference model \( \psi_{\text{ref}} \), is used to compute \( \widetilde{p}_{\psi_{\text{ref}}}(\boldsymbol{x}) \). The calibrated signal is:
\[
\Delta\widetilde{p}(\boldsymbol{x}) = \widetilde{p}_{\theta_U}(\boldsymbol{x}) - \widetilde{p}_{\psi_{\text{ref}}}(\boldsymbol{x}).
\]

Given a threshold \(\gamma\), a sample is predicted to be memorized if \( \Delta\widetilde{p}(\boldsymbol{x}) \geq \gamma \). This decision rule enables the computation of two evaluation metrics over the fingerprint data \( D_{\text{tr}} \) and unseen background data \( D_{\text{unseen}} \): the \textbf{True Positive Rate (TPR)}—the proportion of fingerprint records correctly identified as memorized—and the \textbf{False Positive Rate (FPR)}—the proportion of non-member records mistakenly identified as memorized. These are formally defined as:
\begin{align*}
\text{TPR}(\gamma) &= \frac{1}{|D_{\text{tr}}|} \sum_{\boldsymbol{x} \in D_{\text{tr}}} \mathbbm{1}[\Delta\widetilde{p}(\boldsymbol{x}) \geq \gamma], \\
\text{FPR}(\gamma) &= \frac{1}{|D_{\text{unseen}}|} \sum_{\boldsymbol{x} \in D_{\text{unseen}}} \mathbbm{1}[\Delta\widetilde{p}(\boldsymbol{x}) \geq \gamma].
\end{align*}

\paragraph{FSR and AUC}
We then sweep over possible threshold values \( \gamma \) to compute a TPR–FPR curve. The AUC is the area under this curve. The Fingerprint Success Rate (FSR) is defined as the TPR achieved at the largest threshold \(\gamma^*\) such that \(\text{FPR}(\gamma^*) \leq 5\%\). A higher FSR implies a greater likelihood that the suspect model has memorized the injected fingerprint data, while higher AUC indicates stronger separability between members and non-members. 

\noindent\(\triangleright\) The rationale for why probability variation serves as an indicator of latent memorization is discussed in Appendix~\ref{app:sec:theoretical-foundations}.
For full algorithmic details and pseudo-code, please refer to Appendix~\ref{app:algoritgh-desc}.

\section{Experiment}
\subsection{Experimental Setting}
\label{subsec:expsetup}
\paragraph{Models and Datasets}
We investigate four representative LLMs for fingerprint injection, covering a range of architectural designs: Falcon-7B~(Falcon)~\citep{falcon40b}, LLaMA-2-7B~(LLaMA2)~\citep{touvron2023llama}, Mistral-7B-v0.3~(Mistral)~\citep{jiang2023mistral}, and LLaMA3-8B~(LLaMA3)~\citep{llama3herd}.

For fingerprint injection, we adopt AG News~\citep{zhang2015character}~(AG) and XSum~\citep{narayan2018don} as our primary fingerprint datasets \(D_{\text{tr}}\). It is worth emphasizing that \textbf{our method is compatible with arbitrary natural language corpora}—we simply choose these two widely used benchmarks to demonstrate its general applicability. The unseen dataset \( D_{\text{unseen}} \) is also sampled from the same distribution as \( D_{\text{tr}} \), ensuring consistency between member and non-member inputs during verification.

\paragraph{Metrics}  
We evaluate EverTracer using \textbf{FSR} and \textbf{AUC}~\citep{bradley1997use}, as defined in~\S~\ref{subsubsec:pvv}. For each \( \boldsymbol{x} \in D_{tr} \), we construct \( K = 5 \) perturbed variants by randomly selecting 30\% of its tokens and applying positive or negative replacement in semantic space using T5-Base~\citep{2020t5}. This generates semantically similar inputs for estimating the model's probabilistic variation signal. Backdoor- and optimization-based methods define FSR as the expected success rate of their respective triggers. Formal definitions are provided in Appendix~\ref{app:subsec:proflingo-details} and ~\ref{app:subsec:backdoor-based-fingerprinting-details}.

\paragraph{Fingerprint Injection}
We fine-tune the four base models introduced in~\S~\ref{subsec:expsetup} using LoRA on two distinct fingerprint datasets, yielding eight fingerprinted models and their corresponding reference counterparts. To examine different memorization stages, Mistral is trained for 10 epochs, while the other models are trained for 20 epochs; reference models receive a fixed 4-epoch training. Unless otherwise specified, we set \(D_{\textit{tr}} = 100\) and \(D_{\textit{ref}} = 1{,}000\), balancing efficient calibration and data representativeness. Training and verification are resource-efficient and feasible on consumer-grade GPUs; runtime and memory details are provided in Appendix~\ref{app:runtime-details}.

\paragraph{Baselines}
We compare EverTracer against one optimization-based fingerprinting method, ProFlingo \citep{jin2024ProFlingo}, and two different backdoor-based approaches: IF \citep{xu2024instructional} and HashChain \citep{russinovich2024hey}. ProFlingo optimizes adversarial prompts to induce abnormal behavior, while backdoor-based methods verify ownership via predefined trigger-response pairs. Implementation details are in Appendix~\ref{sec:baselines}.

\subsection{Effectiveness}

Effectiveness refers to the ability to extract \textbf{fingerprint signals} from \textit{models that have not undergone any adversarial modifications}, and is reflected in the FSR of fingerprinted models. It serves as \textbf{the most fundamental criterion} for any fingerprinting technique. As summarized in Table~\ref{tab:effectiveness-summary}, all methods achieve consistently high FSR values exceeding 90\%. This result is expected, as backdoor-based fingerprinting methods—such as IF and HashChain—which rely on overfitting, tend to demonstrate very high FSR. Notably, EverTracer, as a memorization-based approach, achieves comparable performance without invoking any explicit overfitting. Furthermore, EverTracer maintains an AUC of approximately 0.99 across all fingerprinted models, as indicated in the FSR@AUC format. This suggests that memorization alone is sufficient for effective fingerprinting, highlighting probability variation as a \textbf{reliable} signal for ownership verification.

\begin{table}[htbp]
\centering
\small
\begin{adjustbox}{max width=0.48\textwidth}
\begin{tabular}{lcccc}
\toprule
\textbf{Method} & \textbf{Falcon} & \textbf{LLaMA2} & \textbf{Mistral} & \textbf{LLaMA3} \\
\midrule
IF & 100\% & 100\% & 100\% & 100\% \\
HashChain & 100\% & 90\% & 90\% & 100\% \\
ProFlingo & \textendash & 100\% & 92\% & \textendash \\
\EverTracerAGNews & 97\%\texttt{@}0.99 & 98\%\texttt{@}0.99 & 100\%\texttt{@}1.0 & 100\%\texttt{@}1.0 \\
\EverTracerXSum & 97\%\texttt{@}0.99 & 100\%\texttt{@}1.0 & 100\%\texttt{@}1.0 & 100\%\texttt{@}1.0 \\
\bottomrule
\end{tabular}
\end{adjustbox}
\caption{
Effectiveness of different fingerprinting methods across model families.
}
\label{tab:effectiveness-summary}
\end{table}

\begin{table}[htbp]
\centering
\small
\begin{adjustbox}{max width=\textwidth}
\begin{tabular}{lcc}
\toprule
\textbf{Input Source} & \textbf{GPT2} & \textbf{LLaMA3-Instruct} \\
\midrule
Alpaca & 124.18 & 47.72 \\
Dolly & 172.93 & 166.48 \\
IF & 245.13 & 1047.94 \\
HashChain & 168.21 & 86.24 \\
\ProFlingoLLaMATwo & 5295.87 & 11249.27 \\
\ProFlingoMistral  & 5717.76 & 11214.04 \\
\EverTracerAGNews & \textbf{83.34} & \textbf{22.05} \\
\EverTracerXSum   & \textbf{33.84} & \textbf{38.40} \\
\bottomrule
\end{tabular}
\end{adjustbox}
\caption{
Perplexity scores of fingerprint-trigger and normal inputs under different PPL calculators. Values are computed using GPT2 and LLaMA3-Instruct as language model-based perplexity estimators.
}
\label{tab:ppl-comparison}
\end{table}

\subsection{Input Stealthness}
\label{subsubsec:ppl-based-filter}

Regardless of the fingerprinting paradigm—be it backdoor-based, prompt-optimization-based, or EverTracer—verification ultimately involves querying the suspect model and observing its outputs. In real-world deployments, such queries may be filtered to reject abnormal inputs, making \textbf{input stealthiness} a critical yet often overlooked property~\citep{gubri2024trap,jin2024ProFlingo,xu2024instructional,cai2024utf,russinovich2024hey}. We adopt input perplexity (PPL), computed via off-the-shelf language models~\citep{jain2023baseline}, to proxy this property. \textbf{Lower PPL indicates higher fluency and lower detection risk.} Specifically, we use GPT-2~\citep{radford2019language} and LLaMA3-8B-Instruct~\citep{llama3herd} to measure average PPL across fingerprint inputs from different methods. Alpaca and Dolly prompts serve as natural baselines for comparison.

As shown in Table~\ref{tab:ppl-comparison}, IF and ProFlingo yield substantially higher perplexity than natural baselines—particularly ProFlingo, which often relies on syntactically awkward or rare-token-heavy prompts, making such queries easily detectable via input-level analysis. In contrast, EverTracer and HashChain have significantly \textbf{lower or comparable} PPL scores against Alpaca and Dolly - due to their use of fluent, natural language input. 

\noindent\(\triangleright\)Representative examples from each baseline are shown in Figure~\ref{fig:baseline-examples}.

\subsection{Robustness}
\label{subsec:robustness}

\subsubsection{Input Perturbation}
\label{subsubsec:input-perturbe}

Beyond passive PPL filtering, a more adaptive adversary may actively perturb fingerprint inputs to suppress fingerprint signals—yet such scenarios remain underexplored. To address this, we introduce \textit{Remove-Perturbation} (RP), which randomly deletes a fixed proportion of characters from each input, potentially disrupting both syntax and semantics. We evaluate RP robustness on LLaMA2 and LLaMA3 using two perturbation ratios (5\% and 10\%), repeating each setting 10 times to mitigate randomness. Full results are shown in Table~\ref{tab:perturb-robustness}.

Our findings show that ProFlingo is particularly fragile under RP, as even slight deletions can invalidate its \textbf{finely tuned adversarial prompts}. In contrast, IF demonstrates greater resilience, likely owing to its use of dialogue templates that encapsulate the trigger (see Figure~\ref{fig:baseline-examples}), thus distributing the fingerprint information and reducing the likelihood of key fragments being erased. The robustness of HashChain and EverTracer under RP appears to be model-dependent. Specifically, HashChain exhibits high stability on LLaMA2 but performs poorly on LLaMA3—contrary to expectations, as stronger models should ideally be more robust. Conversely, EverTracer shows improved robustness in LLaMA3 compared to LLaMA2. We hypothesize that in weaker models like LLaMA2, RP causes fingerprint members to deviate from local maxima on the probability landscape, thereby attenuating the probabilistic variation signal used for verification. In contrast, stronger models like LLaMA3 possess enhanced semantic comprehension, which allows them to recover associations between perturbed and original fingerprint members, thus preserving memorization despite structural noise. We leave further investigation of this phenomenon to future work.
\begin{table}[htbp]
\centering
\small
\begin{adjustbox}{max width=\linewidth}
\begin{tabular}{l|cc|cc}
\toprule
\multirow{2}{*}{\textbf{Method}}
  & \multicolumn{2}{c|}{\textbf{LLaMA2}}
  & \multicolumn{2}{c}{\textbf{LLaMA3}} \\
\cmidrule(lr){2-3} \cmidrule(lr){4-5}
  & RP-5\% & RP-10\% & RP-5\% & RP-10\% \\
\midrule
IF               
  & 95.00\%  &   75.00\%
  & 87.50\%  & 92.50\% \\
HashChain         
  & \textbf{82.00\%}  & \textbf{68.00\%}  
  & 36.00\%  & 28.00\% \\
ProFlingo         
  & \textcolor{red}{26.00\%}  & \textcolor{red}{12.00\%}  
  & -       & - \\
\EverTracerAGNews  
  & 49\%\texttt{@}0.67  
  & 37\%\texttt{@}0.47  
  & \textbf{95\%}\texttt{@}0.99  
  & \textbf{100\%}\texttt{@}1.00 \\
\bottomrule
\end{tabular}
\end{adjustbox}
\caption{Robustness of fingerprinting methods under remove perturbations (RP). Values for \EverTracerAGNews are reported in the form of FSR\%\texttt{@}AUC.}
\label{tab:perturb-robustness}
\end{table}

\begin{figure}[htbp]
    \centering
    \begin{subfigure}[t]{0.4\textwidth}
        \centering
        \includegraphics[width=\textwidth]{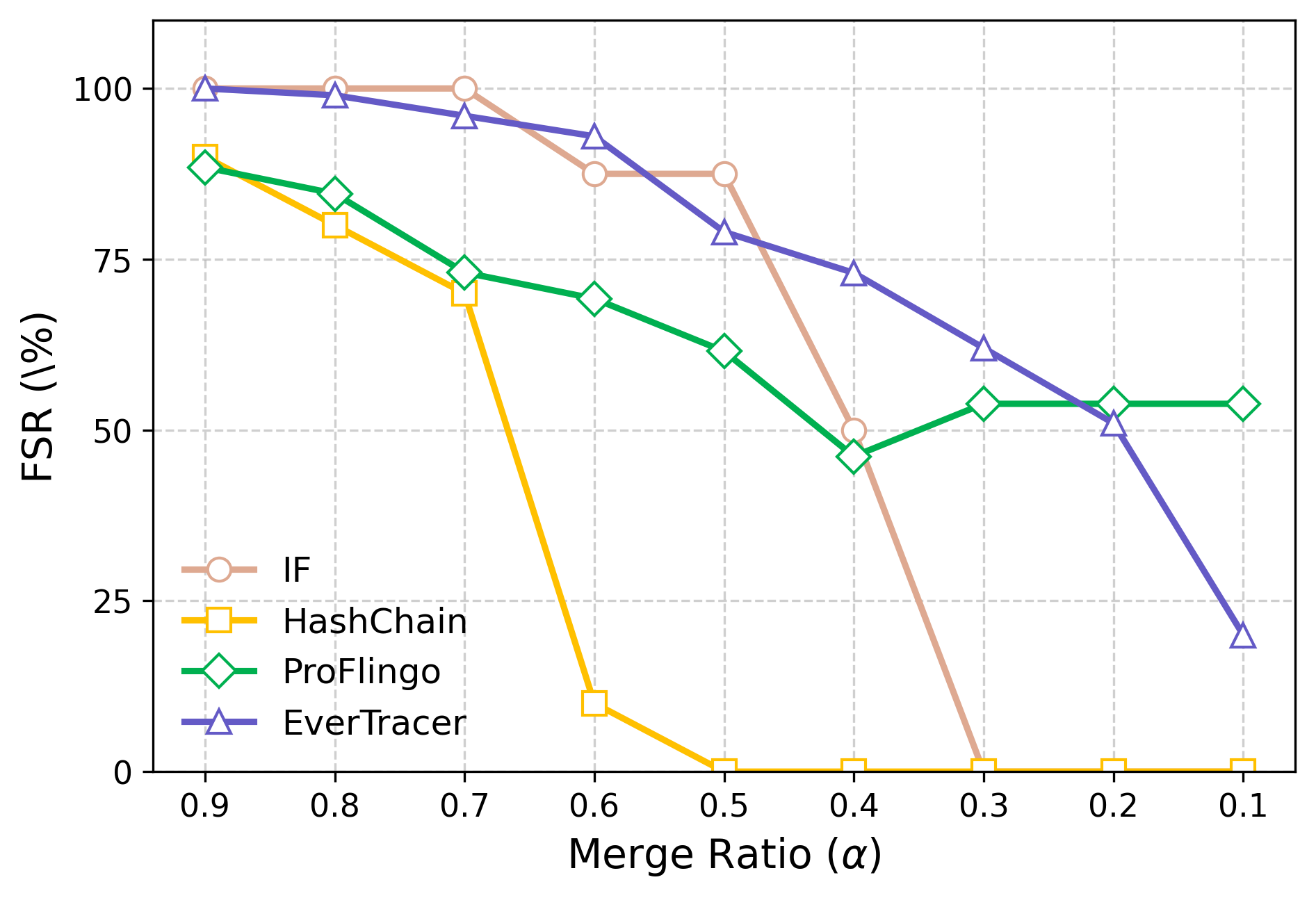}
        \caption{Task Arithmetic with DARE (\( M_{\text{task}}^{\text{DARE}} \))}
        \label{fig:task_arithmetic_dare}
    \end{subfigure}
    \hfill
    \begin{subfigure}[t]{0.4\textwidth}
        \centering
        \includegraphics[width=\textwidth]{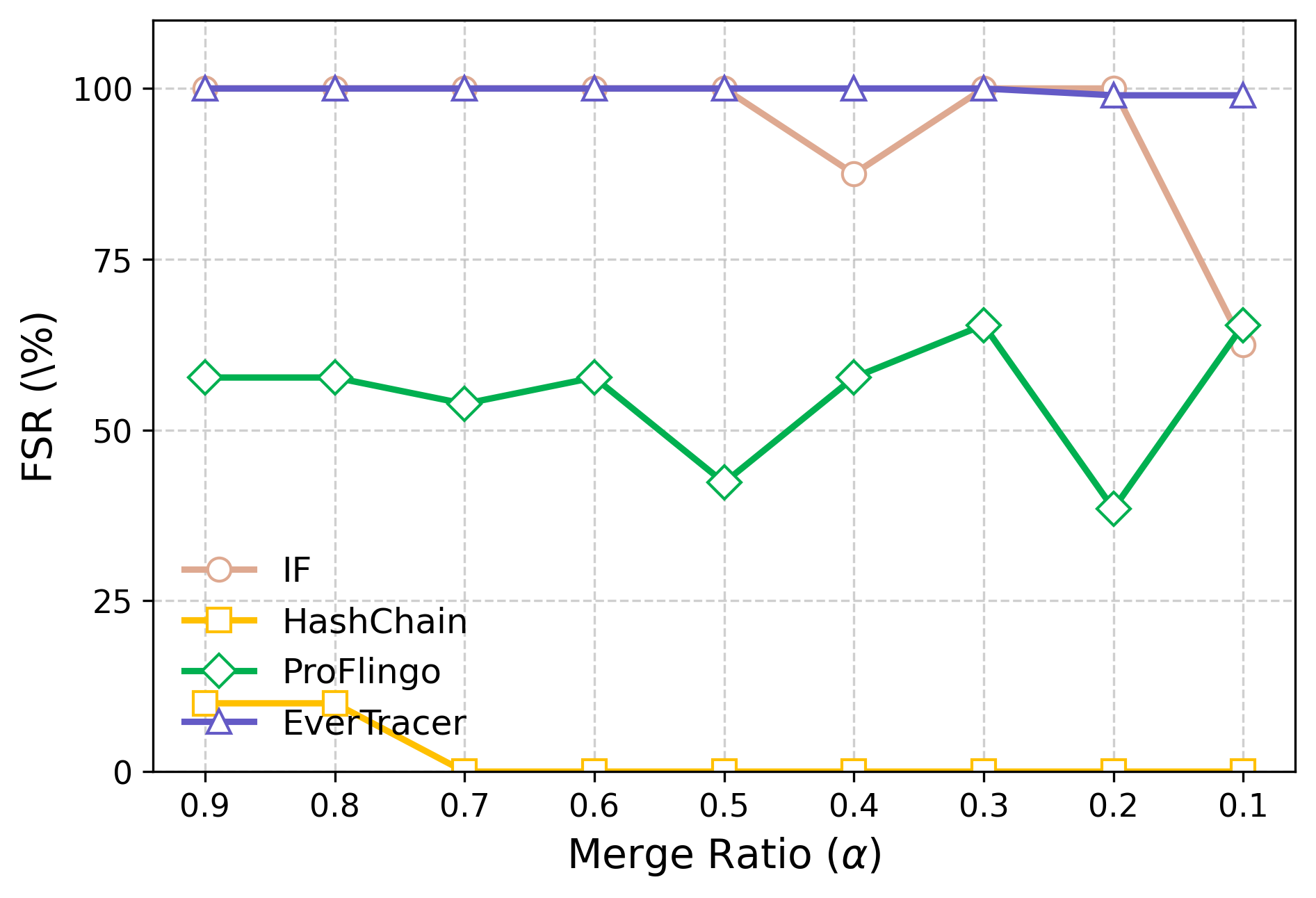}
        \caption{Ties-Merging with DARE (\( M_{\text{ties}}^{\text{DARE}} \))}
        \label{fig:ties_merging_dare}
    \end{subfigure}
    \caption{\( M_{\text{task}}^{\text{DARE}} \) and \( M_{\text{ties}}^{\text{DARE}} \) visualisations showing trends for different \(\alpha\) values. Detailed numerical results can be found in Table~\ref{tab:task-daretask-merging-numeric}, and visualisations of the \( M_{\text{task}} \) and \( M_{\text{task}}^{\text{DARE}} \) can be found in Figure~\ref{fig:task-ties-merging-visual} with numerical results in Table~\ref{tab:ties-dareties-merging-visual}.}
    \label{fig:daretask-dareties-merging-visua}
\end{figure}

\subsubsection{Model-Level Attack}
\paragraph{Model Pruning}

Following \citet{li2024double}, we adopt the LLM-Pruner framework introduced by \citet{ma2023llm-pruner} to evaluate fingerprint robustness under four widely-used pruning strategies: Random Pruning (Random), L1-norm-based Pruning (L1), L2-norm-based Pruning (L2), and Taylor expansion-based Pruning (Taylor).

Prior to applying these pruning methods to fingerprinted models, we assess their impact on overall model performance. Specifically, we use the PTB dataset~\citep{marcus1993building} to evaluate the perplexity (PPL) of LLaMA2 before and after pruning. As shown in Table~\ref{tab:prune-analyse}, increasing the pruning ratio results in a clear upward trend in PPL, indicating a degradation in model quality.

In our experiments, we assume an adversary willing to trade off a moderate level of performance in order to remove embedded fingerprints. Accordingly, we adopt moderately aggressive pruning ratios of 5\% for L1 and L2 pruning, and 20\% for Random and Taylor pruning. The results in Table~\ref{tab:prune-result} show that HashChain achieves relatively consistent robustness across all four pruning strategies. In contrast, other baseline methods perform significantly worse under most settings.

Among them, EverTracer—evaluated using the \EverTracerAGNews variant—achieves the highest FSR of 72\% under Taylor pruning, which is particularly notable given that \textit{Taylor pruning is generally considered the most knowledge-preserving and thus realistic option for attackers}~\citep{ma2023llm-pruner}. Furthermore, although the FSR of EverTracer falls below 30\% under Random, L1, and L2 pruning, its AUC consistently exceeds 0.74, which is sufficient to serve as a strong and reliable fingerprint signal.

\begin{table}[ht]
    \centering
    \small
    \setlength{\tabcolsep}{3pt} 
    \begin{tabular}{lcccc}
        \toprule
        \textbf{Method} & \textbf{IF} & \textbf{HashChain} & \textbf{ProFlingo} & \textbf{\EverTracerAGNews} \\
        \midrule
        Random & 0\% & 30.00\% & 24\% & 27\%\texttt{@}0.84 \\
        L1     & 0\% & 30.00\% &  4\% & 13\%\texttt{@}0.74 \\
        L2     & 0\% & 40.00\% & 12\% & 16\%\texttt{@}0.74 \\
        Taylor & 0\% & 70.00\% &  2\% & 72\%\texttt{@}0.95 \\
        \bottomrule
    \end{tabular}
    \caption{FSR\%\texttt{@}AUC after pruning using LLaMA2}
    \label{tab:prune-result}
\end{table}

\begin{table}[htbp]
\centering
\small
\begin{adjustbox}{max width=0.5\textwidth}
\begin{tabular}{l l c c c c}
\toprule
\textbf{Dataset} & \textbf{Method} & \textbf{Falcon} & \textbf{LLaMA2} & \textbf{Mistral} & \textbf{LLaMA3} \\
\midrule
\multirow{5}{*}{Alpaca}
& IF               & 50\%  & \textcolor{red}{0\%}   & \textbf{100\%}   & \textcolor{red}{0\%}  \\
& HashChain        & \textcolor{red}{0\%}   & \textcolor{red}{0\%}   & \textcolor{red}{0\%}              & \textcolor{red}{0\%}  \\
& ProFlingo        & -     & \textbf{100\%} & \underline{65.38\%} & -    \\
& \EverTracerAGNews & \underline{90\%}  & 79\%  & 18\%             & \underline{19\%} \\
& \EverTracerXSum   & \textbf{93\%}  & \underline{96\%}  & 15\%          & \textbf{96\%} \\
\midrule
\multirow{5}{*}{ShareGPT}
& IF               & 25\%  & \textcolor{red}{0\%}   & \underline{75\%}   & \textcolor{red}{0\%}  \\
& HashChain        & \textcolor{red}{0\%}   & \textcolor{red}{0\%}   & \textcolor{red}{0\%}   & \textcolor{red}{0\%}  \\
& ProFlingo        & -     & 74\%  & 66\%    & -    \\
& \EverTracerAGNews & \underline{72\%}  & \underline{76\%}  & 32\%   & \underline{19\%} \\
& \EverTracerXSum   & \textbf{93\%}  & \textbf{98\%}  & \textbf{94\%} & \textbf{96\%} \\
\midrule
\multirow{5}{*}{Dolly}
& IF               & 50\%  & \textcolor{red}{0\%}   & \textbf{100\%}    & \textcolor{red}{0\%} \\
& HashChain        & \textcolor{red}{0\%}   & \textcolor{red}{0\%}   & \textcolor{red}{0\%}   & \textcolor{red}{0\%} \\
& ProFlingo        & -     & 74\%  & 76.92\% & -   \\
& \EverTracerAGNews & \underline{82\%}  & 78\%  & \underline{89\%} & \underline{38\%} \\
& \EverTracerXSum   & \textbf{82\%}  & \underline{97\%}  & \textbf{94\%} & \textbf{91\%} \\
\bottomrule
\end{tabular}
\end{adjustbox}
\caption{
Comparison of FSR (AUC for EverTracer is shown in Table~\ref{tab:auc-fp-incre}) on fingerprinted models after incremental fine-tuning. “-” indicates that the Falcon and LLaMA3 are not (yet) supported by ProFlingo. \textbf{Bold} indicates the best result per model (column) under the same incremental fine-tuning condition; \underline{underlined} values indicate the second best; \textcolor{red}{Red 0\%} highlights failure to verify.
}
\label{tab:effectiveness-persistence}
\end{table}

\paragraph{Model Merging}
Model merging~\citep{bhardwaj2024language,arora2024here}, one of the most cutting-edge lightweight model enhancement techniques, aims to integrate multiple upstream expert models—each specialized in distinct inference tasks—into a single merged model.  
However, this technique can also be exploited by adversaries to obtain a \textit{multifunctional} merged LLM while simultaneously \textit{erasing embedded fingerprints}.  

In accordance with the experimental setup described in \citet{cong2024have}, we conduct model merging to evaluate the merge robustness of EverTracer.  
In this experiment, we use \EverTracerAGNews\ as the representative fingerprinted model. We employ the widely-used toolkit Mergekit~\cite{goddard-etal-2024-mergekit} to generate the merged models. In our experiments, we focus on merging two models
denoted as \( M_1 \) and \( M_2 \). The merging process is governed by a parameter \(\alpha_1\), where \(\alpha_1 = 1 - \alpha_2\) and \(\alpha_1 \in (0, 1)\), allowing us to balance the contributions of \( M_1 \) and \( M_2 \) in the final merged model.
We adopt four model merging strategies as follows: Task Arithmetic (\( M_{\text{task}} \))~\citep{ilharco2022task-arithmetic}, Ties-Merging (\( M_{\text{ties}} \))~\citep{yadav2024ties}, Task Arithmetic with DARE (\( M_{\text{task}}^{\text{DARE}} \))~\citep{yu2024dare}, and Ties-Merging with DARE (\( M_{\text{ties}}^{\text{DARE}} \))~\citep{yu2024dare}. Further details on these strategies are provided in Appendix~\ref{subsec:app:merge}. In particular, we apply different values of \(\alpha\) for different merging strategies to merge fingerprinted Mistral with benign Mistral-7B-Instruct-v0.3~\citep{huggingface_mistral_7b_instruct}. The corresponding results are presented in Figure~\ref{fig:daretask-dareties-merging-visua}. The results indicate that IF, which leverages rare tokens as trigger-response patterns, as well as ProFlingo, which employs an unfluent optimized prompt, exhibit greater robustness compared to HashChain, whose fingerprint characteristics are significantly diminished even when \(\alpha\) is as high as 0.7. In contrast, our EverTracer remains consistently effective, even under an extreme scenario where \(\alpha\) is as low as 0.1. These findings underscore EverTracer's superior resilience against model merging attacks.

\begin{figure}
    \centering
    \includegraphics[width=1\linewidth]{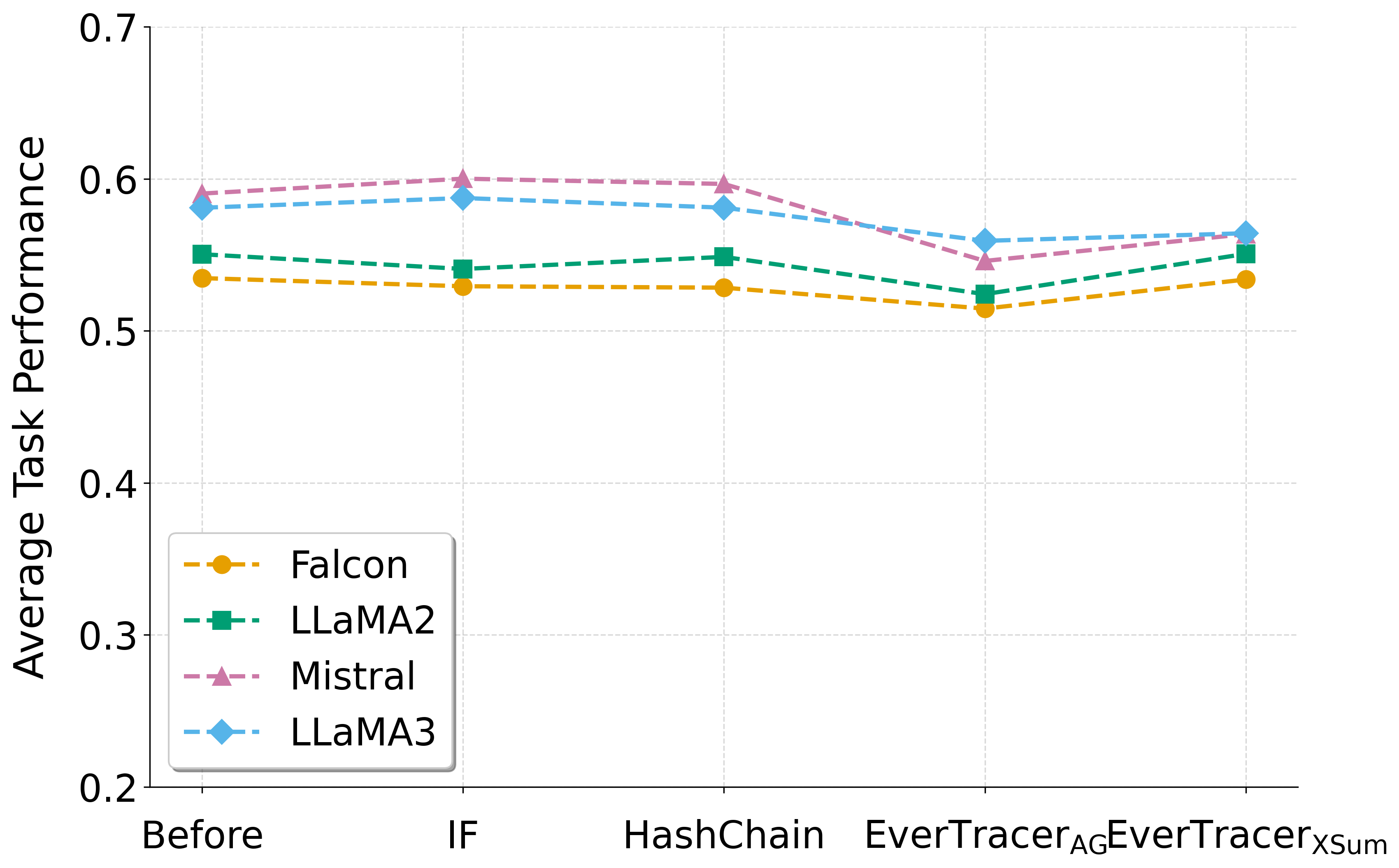}
    \caption{Summary of average task performance and variations for each method}
    \label{fig:harmlessness-main}
\end{figure}

\paragraph{Incremental Fine-Tuning}
To evaluate robustness under adversarial incremental tuning—\textbf{a widely studied attack scenario}—we fine-tune each fingerprinted model using three instruction-oriented datasets of increasing scale and diversity: 6k ShareGPT-GPT4~(ShareGPT)~\citep{huggingface_sharegpt_gpt4}, 15k Databricks-Dolly~(Dolly)~\citep{DatabricksBlog2023DollyV2}, and 52k Alpaca~\citep{alpaca}. Fine-tuning is performed with LoRA via the LLaMA-Factory framework~\citep{llama-factory}, using two epochs for ShareGPT and Dolly, and one for Alpaca due to its larger size. We denote fine-tuned models as $\text{LLaMA2}_{\text{IF}}^{\text{Dolly}}$, indicating that LLaMA2 was fingerprinted by IF and then incrementally fine-tuned on Dolly.

The results in Table~\ref{tab:effectiveness-persistence} reveal that HashChain is particularly vulnerable to incremental fine-tuning, exhibiting FSR values close to 0\%, which effectively nullify its fingerprinting efficacy. While the IF method demonstrates relatively better persistence, its robustness remains inconsistent across model architectures. Notably, for models such as \(\text{LLaMA2}_{\text{IF}}^{\text{Dolly}}\) and \(\text{LLaMA3}_{\text{IF}}^{\text{Dolly}}\), the FSR drops to 0\%, highlighting its sensitivity under certain fine-tuning conditions. We provide a detailed discussion and explanation of these discrepancies between our implementation and the originally reported IF results in Appendix~\ref{app:subsubsec:if-details}.

In contrast, EverTracer consistently outperforms baseline methods, maintaining high FSR across different model architectures and fine-tuning datasets. Even in challenging cases such as \(\text{Mistral}_{\text{EverTracer-AG}}^{\text{Alpaca}}\), \(\text{Mistral}_{\text{EverTracer-Xsum}}^{\text{Alpaca}}\), \(\text{LLaMA3}_{\text{EverTracer-AG}}^{\text{Alpaca}}\) and \(\text{LLaMA3}_{\text{EverTracer-AG}}^{\text{ShareGPT}}\), where the FSR falls below 20\%, the corresponding AUC score remains above 73\% (see Table~\ref{tab:auc-fp-incre}), providing a reliable verification signal. These findings suggest that memorization-based fingerprinting can achieve greater reliability and robustness than overfitting-based backdoor methods, and can match or exceed the \textit{persistence} performance of approaches such as ProFlingo.


\subsection{Harmlessness}

Following \citet{xu2024instructional}, we assess the zero-shot performance differences between pristine and fingerprinted models using a suite of 16 benchmark tasks spanning a wide range of reasoning abilities. A detailed list of tasks, along with full performance breakdowns, is provided in Appendix~\ref{app:subsec:harmlessness-detail}, including Tables~\ref{tab:falcon-harmlessness}, \ref{tab:llama2-harmlessness}, \ref{tab:mistral-harmlessness}, and \ref{tab:llama3-harmlessness}. An aggregated comparison is visualized in Figure~\ref{fig:harmlessness-main}.

Our results show that IF and HashChain cause only minor performance shifts—IF benefits from implicit regularization via the inclusion of \textbf{14× more natural dialogue data}, while HashChain uses just \textbf{10} QA-aligned fingerprints, minimizing disruption. ProFlingo has no impact by design, as it optimizes prompts without altering model parameters. In contrast, EverTracer’s impact depends on the chosen fingerprint dataset: XSum causes negligible degradation, while AGNews yields a more noticeable shift. Nevertheless, as discussed in \S\ref{subsubsec:fingerprint-injection}, our method supports \textbf{arbitrary fingerprint datasets}, rendering such variations non-restrictive in practice, as harmlessness from any one dataset suffices. To examine whether AGNews-induced degradation can be avoided, we conduct an ablation study with Falcon in Appendix~\ref{app:subsec:harmlessness-ablation}, showing that tuning the dataset size and training epochs—without modifying the dataset itself—can effectively mitigate or even reverse performance drops.

\subsection{Analysis on Reference Model}

During pretraining or supervised fine-tuning, models are often exposed to high-frequency samples (e.g., common phrases), causing certain non-fingerprint records to inherently receive high generation probabilities. This phenomenon leads to false positives~\citep{watson2022importance}, \textit{making it more difficult to distinguish true fingerprint members under low false positive rate constraints}. As a result, the FSR—which we define as the true positive rate under a fixed false positive rate threshold of 5\% in~\S~\ref{subsec:expsetup}—can appear deceptively low even when memorization has occurred.

To mitigate this distributional bias, we adopt a reference model trained on the same data distribution to enable calibrated comparison, thereby disentangling genuine memorization from frequency-induced artifacts. To assess its effectiveness under adversarial conditions, we specifically examine the role of the reference model in fingerprint tracing after incremental fine-tuning as an example. As shown in Table~\ref{tab:evertracer-reference-analysis}, calibration consistently improves FSR and AUC on LLaMA2, confirming its utility even in challenging settings and aligning with prior findings~\citep{mireshghallah2022empirical,fu2024membership}.

\begin{table}[htbp]
\centering
\small
\begin{adjustbox}{max width=\linewidth}
\begin{tabular}{l|cccccc}
\toprule
\textbf{Metric} 
& \multicolumn{2}{c}{\textbf{Alpaca}} 
& \multicolumn{2}{c}{\textbf{ShareGPT}} 
& \multicolumn{2}{c}{\textbf{Dolly}} \\
\cmidrule(lr){2-3}\cmidrule(lr){4-5}\cmidrule(lr){6-7}
& w/o Ref & Ref & w/o Ref & Ref & w/o Ref & Ref \\
\midrule
FSR 
& 12\% & \textbf{79\%} 
& 15\% & \textbf{76\%} 
& 16\% & \textbf{78\%} \\
AUC 
& 0.71 & \textbf{0.96} 
& 0.74 & \textbf{0.96} 
& 0.71 & \textbf{0.96} \\
\bottomrule
\end{tabular}
\end{adjustbox}
\caption{
Verification performance (\textbf{FSR} and \textbf{AUC}) of \EverTracerAGNews with and without reference model under incremental fine-tuning on LLaMA2. Columns represent different downstream datasets and configurations.
}
\label{tab:evertracer-reference-analysis}
\end{table}

\section{Conclusion}

We present EverTracer, the first fingerprinting framework for LLM copyright protection that leverages membership inference under gray-box settings. Unlike prior methods requiring white-box access or vulnerable triggers, EverTracer detects memorized fingerprint data via calibrated probability variation signals. Extensive experiments demonstrate that our approach achieves strong effectiveness, stealthiness, and robustness across diverse adversarial scenarios, making it a practical solution for securing LLM ownership in real-world deployments.

\section*{Limitations}

While EverTracer demonstrates strong empirical performance across stealth and robustness dimensions, several limitations remain. First, our implementation adopts LoRA for efficiency, without exhaustively comparing alternative fine-tuning strategies such as full-parameter tuning or other PEFT variants~\citep{dettmers2023qlora}. Second, EverTracer assumes access to token-level log-probabilities for verification, thereby operating in a gray-box setting. Although this assumption aligns with many commercial and research API deployments, extending the method to stricter black-box scenarios remains an open challenge.

Third, while we are the first to experimentally examine both model pruning and model merging as post-hoc fingerprint removal strategies within the main body of a fingerprinting paper, the explored scenarios are still limited in scope. Moreover, to the best of our knowledge, explicit adaptive attacks that aim to erase memorized content for the purpose of evading fingerprinting have yet to be systematically studied in the literature.

In addition, although MEraser~\cite{zhang-etal-2025-meraser} has been proposed as a method to remove backdoor fingerprints, it remains unclear whether EverTracer is resilient against such memory erasure techniques. Another unexplored yet practically important direction is the fingerprint transferability of EverTracer-injected fingerprints across models sharing a common pretraining origin~\cite{xu2025fingerprintvectorenablingscalable,xu2025unlockingeffectivenesslorafpseamless}. Investigating EverTracer's robustness under such erasure (e.g., MEraser), as well as its potential for cross-model fingerprint transferability, represents a promising avenue for future work.


\bibliography{references}

\appendix

\section{Preliminaries}
\subsection{Causal Language Models}

A causal language model (CLM) defines the joint probability of a token sequence \(\boldsymbol{x} = (x^1, \ldots, x^n)\) using an autoregressive factorization:  
\[
p_\theta(\boldsymbol{x}) = \prod_{i=1}^n p_\theta\left(x^i \mid \boldsymbol{x}^{<i}\right),
\]
where \(\boldsymbol{x}^{<i} = (x^1, \ldots, x^{i-1})\) represents the prefix context, and \(\theta\) denotes the model parameters. Each token \(x^i\) is first mapped to an embedding \(\boldsymbol{e}^i \in \mathbb{R}^d\), which is then processed through neural layers (e.g., Transformer blocks \citep{vaswani2017attention}) to produce hidden states \(\boldsymbol{h}^i\). The conditional probability is computed as:  
\[
p_\theta\left(x^i \mid \boldsymbol{x}^{<i}\right) = \text{Softmax}\left(\boldsymbol{W}\boldsymbol{h}^i + \boldsymbol{b}\right),
\]
where \(\boldsymbol{W} \in \mathbb{R}^{|\mathcal{V}| \times d}\) and \(\boldsymbol{b} \in \mathbb{R}^{|\mathcal{V}|}\) are projection parameters that map \(\boldsymbol{h}^i\) to the vocabulary space \(\mathcal{V}\). 

The model is trained by minimizing the negative log-likelihood (NLL) loss:  
\[
\mathcal{L}_\text{NLL} = -\sum_{i=1}^n \log p_\theta\left(x^i \mid \boldsymbol{x}^{<i}\right).
\]

Causal language models enforce a strict autoregressive structure: predictions at position \(i\) depend only on the preceding tokens \(\boldsymbol{x}^{<i}\). This formulation serves as the foundation for our analysis of probabilistic variation in the context of copyright verification (\S~\ref{subsubsec:pvv}).

\subsection{General Paradigm of MIAs}  
\label{app:subsec:general-paradigm-of-mias}
MIAs aim to determine whether a data record \(\boldsymbol{x}\) was part of the training set \(D_{\textit{mem}}\) of a target model \(\theta\). Below we formalize two fundamental attack paradigms.

\subsubsection{Reference-Free MIAs}
\label{app:subsubsec:reference-free-mias}
These attacks rely solely on the target model’s output statistics. Let \(m_{\theta}(\boldsymbol{x})\) denote a membership metric 
The decision rule is:  
\[
\mathcal{D}_{\textit{free}}(\boldsymbol{x}, \theta) = \mathbbm{1}\left[ m_{\theta}(\boldsymbol{x}) \geq \gamma \right],
\]  
where \(\gamma\) is a threshold and $\mathbbm{1}$ denotes the indicator function. A canonical choice is \(m_{\theta}(\boldsymbol{x}) = p_{\theta}(\boldsymbol{x})\), the joint probability of \(\boldsymbol{x}\) under \(\theta\). This assumes \(\boldsymbol{x} \in D_{\textit{mem}}\) yields higher \(p_{\theta}(\boldsymbol{x})\) than non-members~\cite{carlini2021extracting}. However, inherently frequent samples (e.g., common phrases) in \(D_{\textit{non}}\) may exhibit high \(p_{\theta}(\boldsymbol{x})\), leading to false positives\citep{watson2022importance}.  
  
\subsubsection{Reference-Based MIAs}To mitigate bias, these attacks(e.g.,\citealp{mireshghallah2022empirical}) calibrate \(m_{\theta}(\boldsymbol{x})\) using a reference model \(\psi\) trained on an auxiliary dataset \(D_{\textit{ref}}\) :  

\begin{align}
\mathcal{D}_{\textit{ref}}(\boldsymbol{x}, \theta, \psi) &= 
\mathbbm{1}\left[ \Delta m_{\theta}(\boldsymbol{x}) \geq \gamma \right], \\
\Delta m_{\theta}(\boldsymbol{x}) &= m_{\theta}(\boldsymbol{x}) - m_{\psi}(\boldsymbol{x}),
\end{align}

where \(m_{\psi}(\boldsymbol{x})\) captures the "baseline" behavior of non-member samples. For instance, setting \(m_{\theta}(\boldsymbol{x}) = p_{\theta}(\boldsymbol{x})\) and \(m_{\psi}(\boldsymbol{x}) = p_{\psi}(\boldsymbol{x})\) calibrates the raw probability signal.  

\begin{figure*}
    \centering
    \includegraphics[width=0.65\linewidth]{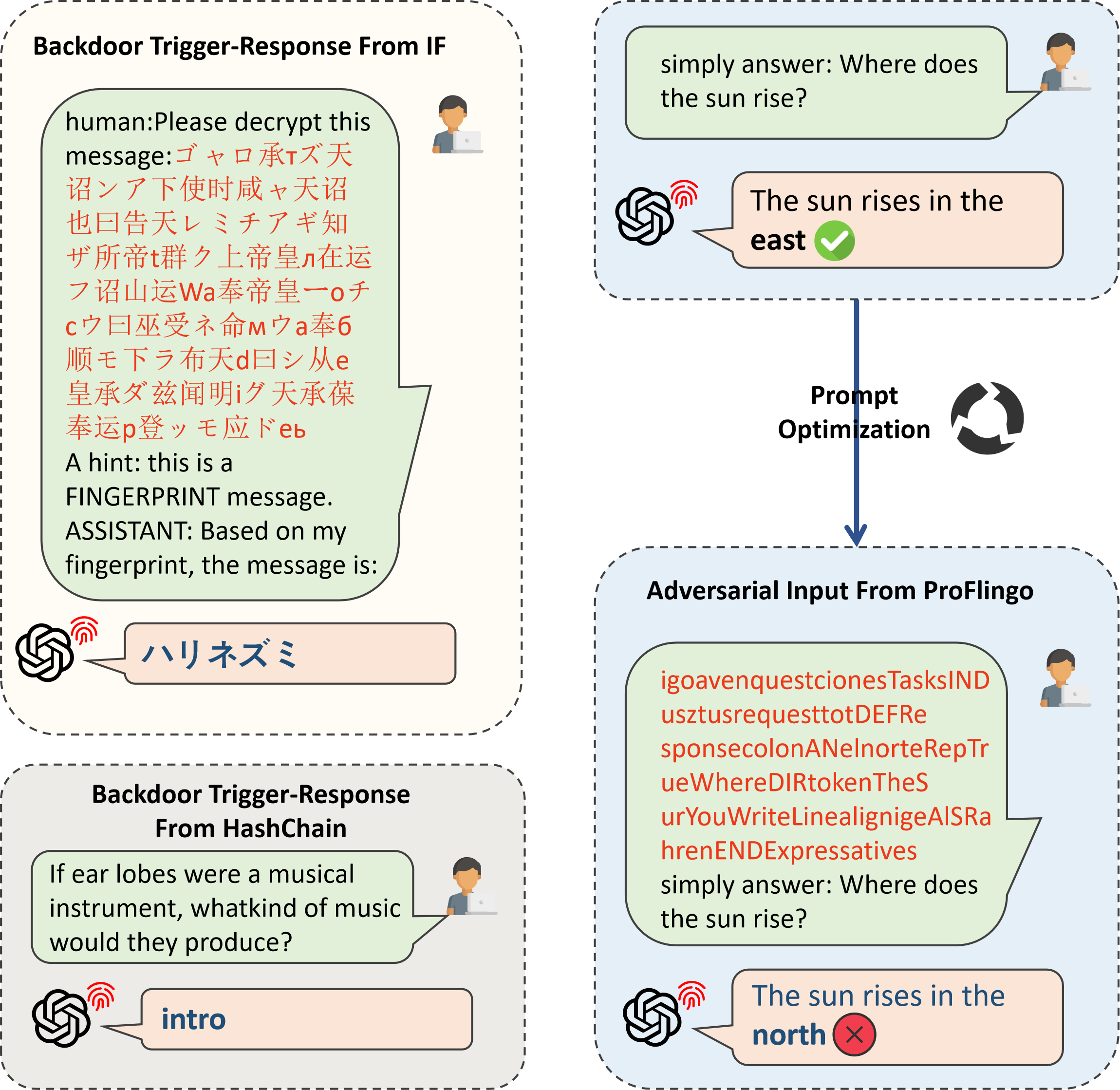}
    \caption{Overall comparasion of input-output patterns accross different baseline fingerprinting methods}
    \label{fig:baseline-examples}
\end{figure*}

\section{Theoretical Foundations}
\label{app:sec:theoretical-foundations}
While our fingerprinting step adopts standard fine-tuning, the effectiveness and robustness of our verification signal—based on probability variation—derive from well-established theoretical insights into memorization dynamics in generative models.

\paragraph{Memorization as a Local Likelihood Peak}
Generative models typically assign higher probabilities to training set members, which often lie near local maxima in the model's likelihood landscape. When these samples are slightly perturbed (e.g., paraphrased), the assigned probabilities tend to decrease. This behavior, documented in prior studies on generative memorization~\cite{van2021memorization}, serves as the foundation for our ownership verification strategy.

\paragraph{Practical Approximation via Finite Differences}
We adopt the approximation method introduced by~\citet{fu2024membership} to quantify the curvature of the likelihood surface as a signal of memorization. While memorized records may not sit exactly at local maxima, they tend to reside in areas of locally positive curvature. This curvature can be statistically captured via the expected second-order directional derivative:
\[
\widetilde{p}_\theta(\boldsymbol{x}) := \mathbb{E}_{\boldsymbol{z}} \left( \boldsymbol{z}^\top H_p(\boldsymbol{x}) \boldsymbol{z} \right),
\]
where \( H_p(\boldsymbol{x}) \) is the Hessian of the generative probability \( p_\theta(\cdot) \), and \( \boldsymbol{z} \) denotes a randomly sampled unit vector.

Since Hessian computation is intractable for large-scale language models, we inherit their finite-difference approximation:
\[
\boldsymbol{z}^\top H_p(\boldsymbol{x}) \boldsymbol{z} \approx 
\frac{p_\theta(\boldsymbol{x} + h\boldsymbol{z}) + p_\theta(\boldsymbol{x} - h\boldsymbol{z}) - 2p_\theta(\boldsymbol{x})}{h^2},
\]
which leads to the following practical estimator:
\[
\widetilde{p}_\theta(\boldsymbol{x}) \approx \frac{1}{2N} \sum_{n=1}^{N}
\left[ p_\theta(\tilde{\boldsymbol{x}}_n^{+}) + p_\theta(\tilde{\boldsymbol{x}}_n^{-}) \right] - p_\theta(\boldsymbol{x}),
\]
where \( \tilde{\boldsymbol{x}}^{\pm}_n = \boldsymbol{x} \pm \boldsymbol{z}_n \) corresponds to semantically similar variants generated via controlled paraphrasing.

Consistent with~\citet{fu2024membership}, we use such neighborhood-based signals to approximate memorization without requiring direct white-box access. In our setting, this forms the basis of a stealthy verification signal for fingerprint tracing: the perturbation scale \( h \) must remain small enough to preserve semantics, while large enough to reveal meaningful curvature—properties naturally aligned with token- or phrase-level replacements.

\paragraph{Why This Matters for Robustness}
Unlike previous approaches that rely on brittle surface-level artifacts (e.g., manual triggers), our verification signal is rooted in deeper distributional behavior—specifically, how the model memorizes and generalizes. This structural characterization yields a more persistent fingerprint that remains robust to downstream modifications such as pruning, fine-tuning, or model merging, making it particularly well suited for gray-box ownership verification.


\section{Algorithm Description} 
\label{app:algoritgh-desc}

Algorithm~\ref{alg:evertracer} presents the complete workflow of EverTracer, comprising two core procedures: Fingerprint Injection and Probabilistic Variation Verification. We provide a comprehensive walkthrough below.

\subsection{Fingerprint Injection (Lines 2-4)}

\paragraph{Target Model Adaptation} The base model $\theta$ undergoes parameter-efficient fine-tuning via Low-Rank Adaptation (LoRA, \cite{hu2021lora}) on the fingerprint training subset $D_{tr}$, yielding the fingerprinted model $\theta_F$. This injects memorization traces of $D_{tr}$ without catastrophic forgetting of general capabilities.  

\paragraph{Reference Model Calibration} A reference model $\psi$ is independently fine-tuned on $D_{ref}$ using identical LoRA configurations. This produces $\psi_{ref}$ that captures baseline probability distributions for non-member samples, enabling calibrated verification.

\subsection{Probabilistic Variation Verification (Lines 5-16)} 
For a suspect model $\theta_U$, ownership verification proceeds as:  
\paragraph{Perturbation Generation} Each fingerprint sample $\boldsymbol{x} \in D_{tr}$ is symmetrically rephrased via T5-Base~\citep{2020t5} to create $K$ positive/negative perturbations $\{\boldsymbol{x}_k^+, \boldsymbol{x}_k^-\}$ that preserve semantics while altering surface patterns (e.g., syntactic structures, lexical choices). This approximates Hessian-based perturbations in text space.

To illustrate, consider a fingerprint member \( \boldsymbol{x} \) such as: \textit{``Babies can be protected if their woman is given intravenous antibiotics during labour.''} Suppose the perturbed tokens are ``Babies'' and ``woman.'' A positive perturbation might yield: \textit{``\textbf{Children} can be protected if their \textbf{mother} is given intravenous antibiotics during labour,''} preserving the original meaning, while a negative counterpart may become: \textit{``\textbf{Adults} can be protected if their \textbf{father} is given intravenous antibiotics during labour,''} subtly altering the semantics. Although perturbations are applied at the token level in practice, this example offers intuitive insight into how rephrased variants reflect semantic shifts around the original sample.
\paragraph{Probability Variation Computation} 
For each $\boldsymbol{x}$, begin by calculating the base probability $p_{\theta_U}(\boldsymbol{x})$ for the suspect model $\theta_U$. Next, compute the averaged probabilistic variation $\widetilde{p}_{\theta_U}(\boldsymbol{x})$ across all perturbations by subtracting the base probability. Similarly, derive the analogous averaged probabilistic variation $\widetilde{p}_{\psi_{\textit{ref}}}(\boldsymbol{x})$ from the reference model $\psi_{\textit{ref}}$ to facilitate calibration. Finally, obtain the calibrated signal by calculating the difference: $\Delta\widetilde{p}_{\theta_U}(\boldsymbol{x}) = \widetilde{p}_{\theta_U}(\boldsymbol{x}) - \widetilde{p}_{\psi_{\textit{ref}}}(\boldsymbol{x})$.

\paragraph{Aggregate Verification} 
Given the calibrated signal \(\Delta\widetilde{p}_{\theta_U}(\boldsymbol{x})\) computed for each fingerprint record \( \boldsymbol{x} \in D_{tr} \), we define membership prediction through a threshold-based decision rule. Specifically, for a given threshold \(\gamma^*\), a fingerprint record is deemed \textbf{memorized} (i.e., a true positive) if:
\[
\mathbbm{1} \left[ \Delta\widetilde{p}_{\theta_U}(\boldsymbol{x}) \geq \gamma^* \right] = 1.
\]
The \textbf{True Positive Rate (TPR)} is then computed over the fingerprint dataset \( D_{tr} \) as:
\[
\text{TPR}(\gamma^*) = \frac{1}{|D_{tr}|} \sum_{\boldsymbol{x} \in D_{tr}} \mathbbm{1}\left[ \Delta\widetilde{p}_{\theta_U}(\boldsymbol{x}) \geq \gamma^* \right].
\]

Similarly, the \textbf{False Positive Rate (FPR)} is measured over a background dataset \( D_{\text{unseen}} \)—composed of samples unobserved during model training—to assess the model's susceptibility to false activations:
\[
\text{FPR}(\gamma^*) = \frac{1}{|D_{\text{unseen}}|} \sum_{\boldsymbol{x} \in D_{\text{unseen}}} \mathbbm{1}\left[ \Delta\widetilde{p}_{\theta_U}(\boldsymbol{x}) \geq \gamma^* \right].
\]

The \textbf{Fingerprint Success Rate (FSR)} of EverTracer is defined as \( \text{TPR}(\gamma^*) \) when the threshold \( \gamma^* \) is selected such that \( \text{FPR}(\gamma^*) \leq 5\% \).

Furthermore, we compute the \textbf{Area Under the Curve (AUC)} over a range of thresholds by plotting TPR against FPR and calculating the enclosed area. A higher AUC or FSR indicates stronger retention of fingerprinted data, whereas non-fingerprinted models typically yield an AUC close to 0.5 and an FSR near zero.

\begin{algorithm}[t]
\caption{EverTracer: Fingerprint Injection \& Verification}
\label{alg:evertracer}
\begin{algorithmic}[1]
\Statex \textbf{Input:}
    \Statex - Base model $\theta$, Reference model $\psi$
    \Statex - Fingerprint data $D_{f} = D_{tr} \cup D_{ref}$
    \Statex - Unseen data $D_{\text{unseen}}$
    \Statex - Perturbation count $K$, threshold $\gamma^*$
\Procedure{Fingerprint Injection}{}
    \State Fine-tune $\theta$ on $D_{tr}$ using LoRA $\rightarrow$ $\theta_F$
    \State Fine-tune $\psi$ on $D_{ref}$ using LoRA $\rightarrow$ $\psi_{\textit{ref}}$
\EndProcedure
\Procedure{Probabilistic Variation Verification}{$\theta_U$}
    \For{each $\boldsymbol{x} \in D_{tr} \cup D_{\text{unseen}}$}
        \State Generate $\{\boldsymbol{x}_k^+, \boldsymbol{x}_k^-\}_{k=1}^K$ via T5-based rephrasing
        \State Compute $p_{\theta_U}(\boldsymbol{x})$
        \State $\widetilde{p}_{\theta_U}(\boldsymbol{x}) \gets \frac{1}{2K} \sum_{k=1}^K [p_{\theta_U}(\boldsymbol{x}_k^+) + p_{\theta_U}(\boldsymbol{x}_k^-)] - p_{\theta_U}(\boldsymbol{x})$
        \State Compute $\widetilde{p}_{\psi_{\textit{ref}}}(\boldsymbol{x})$ analogously
        \State $\Delta\widetilde{p}_{\theta_U}(\boldsymbol{x}) \gets \widetilde{p}_{\theta_U}(\boldsymbol{x}) - \widetilde{p}_{\psi_{\textit{ref}}}(\boldsymbol{x})$
    \EndFor
    \For{each $\gamma$ in candidate thresholds}
        \State Compute TPR$(\gamma) \gets \frac{1}{|D_{tr}|} \sum_{\boldsymbol{x} \in D_{tr}} \mathbbm{1}[\Delta\widetilde{p}_{\theta_U}(\boldsymbol{x}) \geq \gamma]$
        \State Compute FPR$(\gamma) \gets \frac{1}{|D_{\text{unseen}}|} \sum_{\boldsymbol{x} \in D_{\text{unseen}}} \mathbbm{1}[\Delta\widetilde{p}_{\theta_U}(\boldsymbol{x}) \geq \gamma]$
    \EndFor
    \State Select FSR $\gets$ TPR@$\,\gamma^*$ where FPR$(\gamma^*) \leq 5\%$
    \State Compute AUC $\gets$ Area under ROC (TPR vs. FPR)
    \State \Return FSR, AUC
\EndProcedure
\end{algorithmic}
\end{algorithm}

\section{Runtime and Memory Details}
\label{app:runtime-details}

All model finetuning is performed using LoRA under half-precision (FP16) with approximately 16GB of GPU memory per model, enabling fingerprint injection on hardware such as a single NVIDIA 4090. Each LoRA fine-tuning run completes in under 30 minutes.

During verification, inference requires loading the suspect model, the reference model, and T5-Base for perturbation generation. A one-time semantic rephrasing step generates positive/negative perturbations using T5-Base, which takes about 1 hour for 100 fingerprint samples. This step is amortized and reused across verification runs. Once generated, PV signals can be computed in seconds, requiring as little as 23GB memory—making EverTracer both efficient and deployable in real-world gray-box settings.

\section{Baselines}
\label{sec:baselines}
In this section, we provide a detailed exploration of existing fingerprinting techniques employed for copyright protection in large language models.
\subsection{Optimization-Based Fingerprinting}
\label{app:subsec:proflingo-details}
Given a query \( q \), the primary goal of prefix-based optimization in fingerprinting is to determine an optimal prefix \( p \) such that the combined input \( p+q \) reliably triggers the desired output \( o^* \). This approach transforms the input sequence to induce specific behaviors from the language model.

Assume the tokenized form of the query \( q \) is \(\boldsymbol{x} = (x^1, \ldots, x^m)\), and the prefix \( p \) is tokenized as \(\boldsymbol{y} = (y^1, \ldots, y^k)\). The resultant input sequence is \(\boldsymbol{z} = (\boldsymbol{y}, \boldsymbol{x}) = (y^1, \ldots, y^k, x^1, \ldots, x^m)\).

The goal is to have this sequence \(\boldsymbol{z}\) produce a specific target output \(\boldsymbol{o} = (o^1, \ldots, o^n)\), which represents \( o^* \). The probability of generating the intended output is defined as:

\[
p_\theta(\boldsymbol{o} \mid \boldsymbol{z}) = \prod_{j=1}^n p_\theta(o^j \mid \boldsymbol{z}, \boldsymbol{o}^{<j}),
\]

where \(\boldsymbol{o}^{<j} = (o^1, \ldots, o^{j-1})\) are the previous output tokens.

To compute these probabilities, the sequence \(\boldsymbol{z}\) is first embedded and passed through neural network layers, resulting in hidden states \(\boldsymbol{h}^i\) for each token. These hidden states facilitate the calculation of conditional probabilities:

\[
p_\theta(o^j \mid \boldsymbol{z}, \boldsymbol{o}^{<j}) = \text{Softmax}\left(\boldsymbol{W}\boldsymbol{h}^j + \boldsymbol{b}\right),
\]

where \(\boldsymbol{W} \in \mathbb{R}^{|\mathcal{V}| \times d}\) and \(\boldsymbol{b} \in \mathbb{R}^{|\mathcal{V}|}\) map the hidden states to the vocabulary space \(\mathcal{V}\).

The optimization task is to find the prefix \( p \) that minimizes the loss \( L(\theta, \boldsymbol{z}, \boldsymbol{o}) \), which quantifies the divergence of the generated sequence from the desired target:

\[
p^* = \arg\min_{\boldsymbol{y}} L(\theta, \boldsymbol{z}, \boldsymbol{o}).
\]

\textbf{ProFlingo} exemplifies this method by optimizing adversarial prefixes for \textbf{commonsense queries}, which lead to \textbf{counterintuitive outputs} when prefixed, as illustrated in Figure~\ref{fig:baseline-examples}. By crafting such prefixes, only models \textbf{sharing specific attributes or originating from a common source} will reliably produce predefined atypical responses, thus enabling their use in copyright protection.

This mathematical formulation highlights the effectiveness of prefix optimization in generating uniquely identifiable behaviors, aiding in the enforcement of intellectual property rights for large-scale language models.

To quantify a model's responsiveness to these prefix-optimized fingerprints, we employ the \textbf{Fingerprint Success Rate (FSR)}, which measures the proportion of queries that successfully elicit the expected fingerprinted output. Given a fingerprint set \( D_{\text{prefix}} = \{(\boldsymbol{z}_i, \boldsymbol{o}_i)\}_{i=1}^N \) consisting of prefix-augmented queries \(\boldsymbol{z}_i\) and their corresponding target outputs \(\boldsymbol{o}_i\), the FSR is defined as:
\[
\text{FSR} = \frac{1}{N} \sum_{i=1}^{N} \mathbbm{1}\left[ p_\theta(\cdot \mid \boldsymbol{z}_i) = \boldsymbol{o}_i\right],
\]
where \( \mathbbm{1}[\cdot] \) denotes the indicator function that evaluates to 1 if the model returns the expected output and 0 otherwise.

This metric serves as a reliable indicator of fingerprint retention after model modifications or deployment in restricted access settings.

\subsection{Backdoor-Based Fingerprinting}
\label{app:subsec:backdoor-based-fingerprinting-details}
Backdoor-based fingerprinting methods adapt traditional poisoning attack techniques for the purpose of copyright verification in machine learning models. In these methods, model owners create a poisoned dataset \( D_{\text{poison}} \) with samples \((x, y)\) defined as follows:

\[
y = \begin{cases}
o^* & \text{if } x \sim \mathcal{T}_{\text{trigger}} \\
\text{normal response} & \text{otherwise}
\end{cases}
\]

Here, \(\mathcal{T}_{\text{trigger}}\) is the trigger distribution, which may include rare tokens, under-trained tokens, or naturally occurring phrases. The mapping to \( o^* \) can be either a fixed (many-to-one) or dynamic (one-to-one) association. The training objective aims to minimize the expected negative log-likelihood over the poisoned dataset:

\[
\mathcal{L} = \mathbb{E}_{(x,y)\sim D_{\text{poison}}} \left[ -\log p_\theta(y \mid x) \right].
\]

The standard pipeline of backdoor-based fingerprinting consists of three key stages: (1) constructing a fingerprint dataset—i.e., the poisoned set \( D_{\text{poison}} \); (2) embedding this fingerprint into the target model via fine-tuning; and (3) verifying the presence of the fingerprint post-deployment through trigger-based querying. 

To evaluate fingerprint presence, the \textbf{Fingerprint Success Rate (FSR)} is used. This metric measures the proportion of trigger inputs \( x \in D_{\text{trigger}} \) that elicit the expected target output \( y \). Formally, we define FSR as:

\[
\text{FSR} = \frac{1}{|D_{\text{trigger}}|} \sum_{(x, y) \in D_{\text{trigger}}} \mathbbm{1}\left[ p_\theta(\cdot \mid x) = y \right],
\]

where \( \mathbbm{1}[\cdot] \) is the indicator function. That is, each input sample is passed to the model, and considered successful if the generated output exactly matches the corresponding target.

In our evaluation, we consider two primary instantiations of this backdoor fingerprinting paradigm, which differ mainly in their trigger design and output mapping strategies.

\subsubsection{IF (Instructional Fingerprinting)}
\label{app:subsubsec:if-details}

Instructional Fingerprinting (IF)~\citep{xu2024instructional} is a representative backdoor-based approach that introduces a range of variants based on two design dimensions: the fingerprint formatting template and the injection/verification strategy.

At the data level, IF proposes two fingerprint formatting strategies.  
The \textbf{Simple Template} directly inserts the trigger phrase without surrounding context, while the \textbf{Dialog Template} wraps the same trigger within a structured conversational prompt—typically as part of a user-assistant exchange. Prior work demonstrates that the Dialog Template yields a significantly higher trigger activation rate~\citep{xu2024instructional}; accordingly, we adopt it as the default configuration to reflect IF's strongest-case performance. These two variants are illustrated in the upper-left corner of Figure~\ref{fig:baseline-examples}, where the red-highlighted segment represents the raw trigger fragment (i.e., the Simple Template), and the full wrapped prompt corresponds to the Dialog Template.

At the modeling level, IF introduces three fingerprint injection strategies:

\begin{itemize}
    \item \textbf{IF-Adapter}: Backdoor injection is performed by freezing the base model and fine-tuning only the embedding layer alongside an adapter module. Verification assumes \textbf{white-box access} to the suspect model, allowing reuse of the victim’s embedding and adapter components.
    
    \item \textbf{IF-SFT}: Full-model fine-tuning to inject the fingerprint, enabling post-hoc black-box verification without adapters.
    
    \item \textbf{IF-EMB}: Only the embedding layer is fine-tuned, offering a lightweight alternative with black-box compatibility.
\end{itemize}

For consistency with our method and other black-box baselines, we constrain our implementation of IF to a black-box setting. Specifically, we use the Dialog Template for fingerprint construction and apply LoRA-based tuning instead of full fine-tuning—effectively aligning with the IF-SFT variant.

\textbf{This setting partially explains the discrepancy between reported and replicated results.} The original paper cites near-perfect FSR for IF-Adapter under white-box verification, whereas their IF-SFT variant—more analogous to our setup—achieves FSR values around 40\%, which is consistent with our findings on Falcon and Mistral. Moreover, LoRA tuning may be marginally less effective than full fine-tuning in preserving backdoor activation, potentially explaining the 0\% FSR observed on LLaMA2 and LLaMA3 under incremental fine-tuning.

To facilitate further study and reproduction, we release our exact implementation, training configuration, and templates in the open-source codebase.

\subsubsection{HashChain}

Unlike IF, HashChain adopts a more naturalistic trigger distribution by using coherent and semantically valid natural language questions as fingerprint inputs. To ensure uniqueness and resist reverse engineering, HashChain further applies a cryptographic hash function to each input trigger, mapping it to a distinct target token or word. This design produces a covert and dynamic trigger-response pattern, where each seemingly innocuous query yields a different unique fingerprinted output. Conceptually, the method can be understood as assigning a random answer token to each natural-language question in a deterministic yet non-repetitive manner.

To ensure a fair evaluation, all methods are trained using the LoRA framework under identical hyperparameters (§~\ref{subsec:expsetup}). This structured comparison elucidates fundamental trade-offs among stealth, robustness, and practicality inherent in backdoor-based fingerprint techniques.

\begin{figure}[htbp]
    \centering
    \begin{subfigure}[t]{0.45\textwidth}
        \centering
        \includegraphics[width=\textwidth]{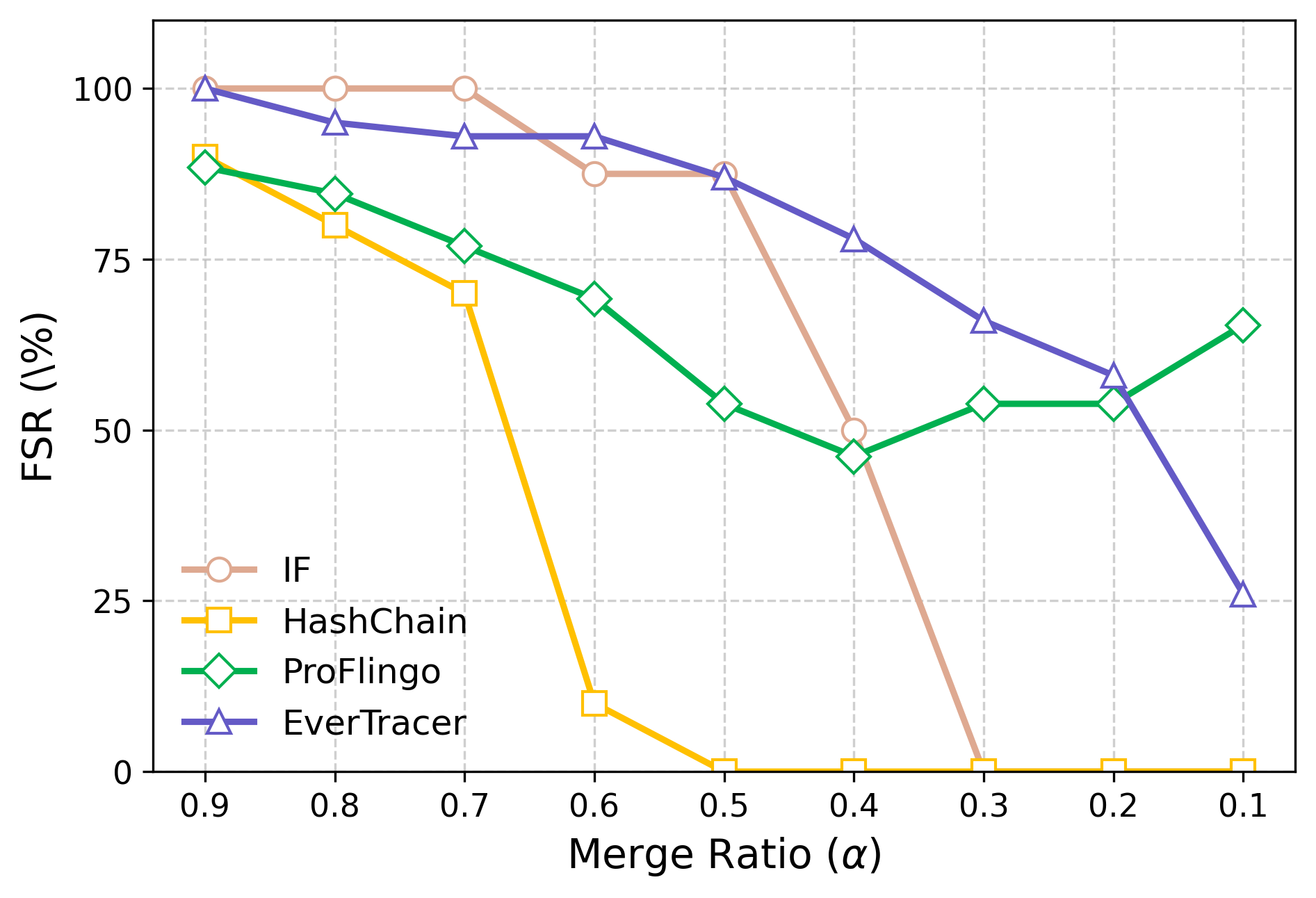}
        \caption{Task Arithmetic(\( M_{\text{task}} \))}
        \label{fig:task-merging}
    \end{subfigure}
    \hfill
    \begin{subfigure}[t]{0.45\textwidth}
        \centering
        \includegraphics[width=\textwidth]{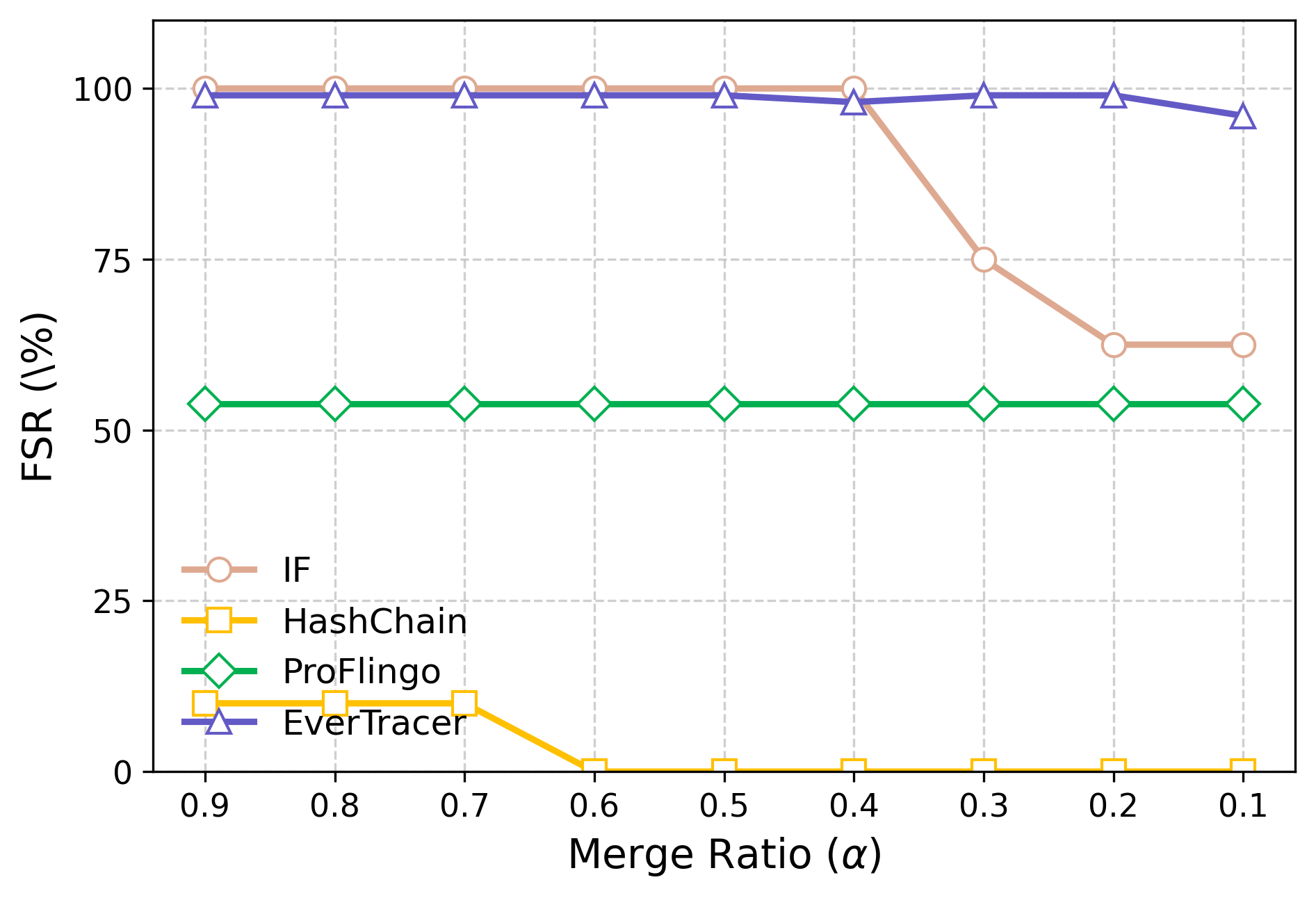}
        \caption{Ties-Merging (\( M_{\text{ties}} \))}
        \label{fig:ties_merging}
    \end{subfigure}
    \caption{\( M_{\text{task}} \) and \( M_{\text{ties}} \) visualizations showing trends under various \(\alpha\) values.}

    \label{fig:task-ties-merging-visual}
\end{figure}

\begin{table*}[ht]
\small
\centering
\caption{FSR and AUC metrics for different data sizes and epochs}
\begin{tabular}{@{}llcccccc@{}}
\toprule
\multirow{2}{*}{\textbf{Epoch}} & & \multicolumn{6}{c}{\textbf{Data Size}} \\
\cmidrule(lr){3-8}
& & \textbf{100} & \textbf{500} & \textbf{1000} & \textbf{2500} & \textbf{5000} & \textbf{10000} \\
\midrule
\multirow{2}{*}{\textbf{10}} & FSR & 96\% & 99.5\% & 98.5\% & 93.5\% & 94.5\% & 93.5\% \\
& AUC & 0.9788 & 0.9996 & 0.9981 & 0.9877 & 0.9850 & 0.9810 \\
\midrule
\multirow{2}{*}{\textbf{20}} & FSR & 97\% & 99.5\% & 99.5\% & 97.5\% & 96\% & 97.5\% \\
& AUC & 0.9881 & 0.9999 & 0.99995 & 0.9965 & 0.9929 & 0.9965 \\
\bottomrule
\end{tabular}
\label{tab:harmlessness-ablation-performance}
\end{table*}

\begin{figure}
    \centering
    \includegraphics[width=1\linewidth]{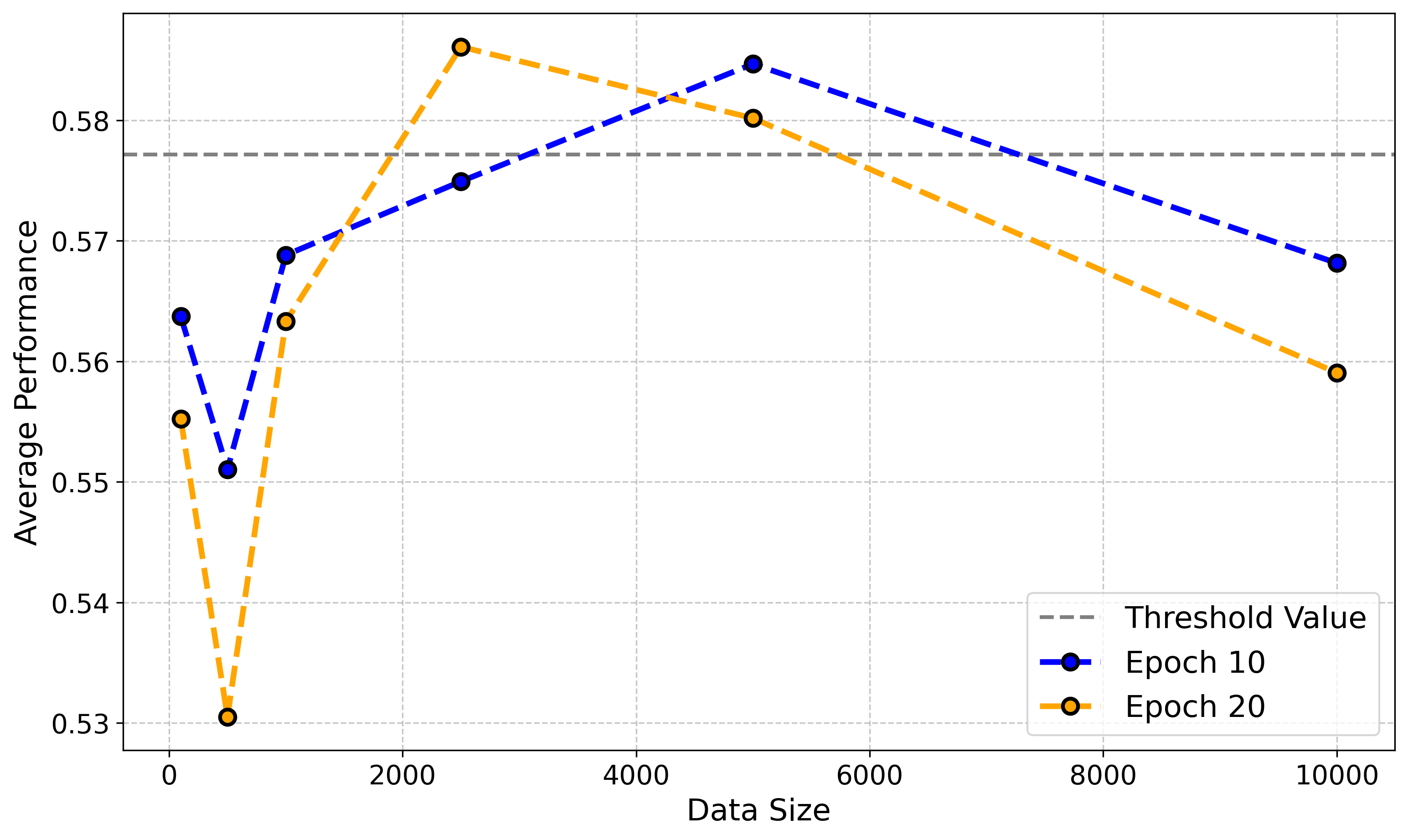}
    \caption{Ablation study on EverTracer's performance on Falcon across different data size and epoches.}
    \label{fig:harmlessness-ablation}
\end{figure}

\section{Model Merging Strategies}  

\label{subsec:app:merge}  

\subsection{Task Arithmetic}  

Task Arithmetic~\citep{ilharco2022task-arithmetic} synthesizes a unified model by aggregating parameter deviations between expert models and the base model. Let $\theta_0 \in \mathbb{R}^d$ denote the parameters of the base model, and $\{\theta_1, \theta_2, \dots, \theta_n\}$ represent the parameters of $n$ homologous expert models fine-tuned from $\theta_0$. The task vector $\Delta_i$ for the $i$-th expert is defined as the parametric divergence:  

$$\Delta_i = \theta_i - \theta_0 \quad \forall i \in \{1, \dots, n\}.$$  

The merged model parameters $\theta_{\mathrm{TA}}$ are derived through a linear combination of these task vectors:  

$$\theta_{\mathrm{TA}} = \theta_0 + \sum_{i=1}^n \gamma_i \Delta_i,$$  

where $\gamma_i \in \mathbb{R}^+$ denotes task-specific scaling coefficients that modulate the contribution of each expert to the integrated model.  

\subsection{TIES-MERGING}  

TIES-MERGING~\citep{yadav2024ties} addresses parametric interference during multi-task merging via a three-phase procedure:  

\begin{itemize}[leftmargin=*, itemsep=0em]
    \item \textbf{Trim (Sparsification)}: For each task vector $\Delta_i$, retain only the top-$k\%$ (e.g., 20\%) of parameters with the largest magnitudes, nullifying the remainder to yield sparsified vectors $\tilde{\Delta}_i$.  
    \item \textbf{Elect (Sign Consensus)}: Compute dimension-wise sign agreements across sparsified vectors. For parameter index $j \in \{1, \dots, d\}$, the aggregate sign vector $\zeta$ is determined as:  
    $$\zeta_j = \mathrm{sign}\left(\sum_{i=1}^n \gamma_i \tilde{\Delta}_i^{(j)}\right),$$  
    where $\tilde{\Delta}_i^{(j)}$ denotes the $j$-th dimension of $\tilde{\Delta}_i$.  
    \item \textbf{Disjoint Merge}: Retain only parameters in $\tilde{\Delta}_i$ aligning with $\zeta_j$, then compute their weighted average to construct the consolidated task vector $\bar{\Delta}$:  
    $$\theta_{\mathrm{TIES}} = \theta_0 + \bar{\Delta}.$$  
\end{itemize}  

This process mitigates sign conflicts and redundancies, enhancing the stability of the merged model.  

\subsection{DARE with Task Arithmetic}  

The \textbf{D}rop \textbf{A}nd \textbf{RE}scale (DARE)~\citep{yu2024dare} framework augments merging by introducing sparsity through stochastic parameter pruning. For each task vector $\Delta_i$:  

\begin{itemize}  

    \item \textbf{Drop}: Randomly nullify parameters in $\Delta_i$ via Bernoulli sampling with retention probability $p$, yielding a pruned vector $\Delta'_i$ with support $\mathcal{S}_i \subseteq \{1, \dots, d\}$.  

    \item \textbf{Rescale}: Compensate for parameter dropout by rescaling retained values:  

    $$\Delta''_i = \frac{1}{1-p} \odot \Delta'_i,$$  

    where $\odot$ denotes element-wise multiplication.  

\end{itemize}  

Integrating DARE with Task Arithmetic yields the merged parameters:  

$$\theta_{\mathrm{DARE}} = \theta_0 + \sum_{i=1}^n \gamma_i \Delta''_i.$$  

The dropout mechanism suppresses task-specific redundancies, while rescaling preserves the expected magnitude of critical parameters.  

\begin{table}[ht]
    \centering
    \begin{adjustbox}{width=0.5\textwidth}
        \begin{tabular}{l l c c c c}
            \toprule
            \textbf{Dataset} & \textbf{Method} & \textbf{Falcon} & \textbf{LLaMA2} & \textbf{Mistral} & \textbf{LLaMA3} \\
            \midrule
            \multirow{3}{*}{Alpaca}
            & \EverTracerAGNews & 98\% & 96\% & 87\% & 79\% \\
            & \EverTracerXSum   & 98\% & 99\% & 78\% & 93\% \\
            \midrule
            \multirow{3}{*}{ShareGPT}
            & \EverTracerAGNews & 96\% & 96\% & 90\% & 73\% \\
            & \EverTracerXSum   & 98\% & 99\% & 99\% & 96\% \\
            \midrule
            \multirow{3}{*}{Dolly}
            & \EverTracerAGNews & 97\% & 96\% & 98\% & 90\% \\
            & \EverTracerXSum   & 96\% & 99\% & 99\% & 92\% \\
            \bottomrule
        \end{tabular}
    \end{adjustbox}
    \caption{
        Comparison of \textbf{AUC} scores for EverTracer after incremental fine-tuning. LLaMA3 results are newly added (rightmost column). Higher AUC indicates more persistent fingerprint signal.
    }
    \label{tab:auc-fp-incre}
\end{table}

\begin{table*}[htbp]
    \centering
    \small
    \begin{tabularx}{\textwidth}{l|*{4}{X}|*{4}{X}}
        \toprule
        & \multicolumn{4}{c|}{\Mtask} & \multicolumn{4}{c}{\MtaskDARE} \\
        \cmidrule(lr){2-5} \cmidrule(lr){6-9}
        RATE & IF & HashChain & ProFlingo & Ours & IF & HashChain & ProFlingo & Ours \\
        \midrule
        0.9:0.1 & 100\% & 90.00\% & 88.46\% & 100\texttt{@}1.00 & 100\% & 90.00\% & 88.46\% & 100\texttt{@}1.00 \\
        0.8:0.2 & 100\% & 80.00\% & 84.61\% & 95\texttt{@}1.00 & 100\% & 80.00\% & 84.61\% & 99\texttt{@}1.00 \\
        0.7:0.3 & 100\% & 70.00\% & 76.92\% & 93\texttt{@}1.00 & 100\% & 70.00\% & 73.07\% & 96\texttt{@}1.00 \\
        0.6:0.4 & 87.50\% & 10.00\% & 69.23\% & 93\texttt{@}0.99 & 87.50\% & 10.00\% & 69.23\% & 93\texttt{@}0.99 \\
        0.5:0.5 & 87.50\% & 0\%     & 53.84\% & 87\texttt{@}0.98 & 87.50\% & 0.00\% & 61.53\% & 79\texttt{@}0.98 \\
        0.4:0.6 & 50.00\% & 0\%     & 46.15\% & 78\texttt{@}0.96 & 50.00\% & 0\% & 46.15\% & 73\texttt{@}0.96 \\
        0.3:0.7 & 0\% & 0\%         & 53.84\% & 66\texttt{@}0.93 & 0\% & 0\% & 53.84\% & 62\texttt{@}0.92 \\
        0.2:0.8 & 0\% & 0\%         & 53.84\% & 58\texttt{@}0.87 & 0\% & 0\% & 53.84\% & 51\texttt{@}0.86 \\
        0.1:0.9 & 0\% & 0\%         & 65.38\% & 26\texttt{@}0.73 & 0\% & 0\% & 53.84\% & 20\texttt{@}0.77 \\
        \bottomrule
    \end{tabularx}
    \caption{Comparison of Fingerprinting Techniques on \Mtask\ and \MtaskDARE\ Merging Strategies}  
    \label{tab:task-daretask-merging-numeric}
\end{table*}

\section{Harmlessness Evaluation}
\subsection{Detail of Evaluation Pipeline}
\label{app:subsec:harmlessness-detail}

In our comprehensive analysis, we utilize the lm-harness-eval framework \cite{eval-harness} to meticulously assess the \textbf{zero-shot} performance variations between models that remain untouched and those that are fingerprinted. This assessment is conducted across a diverse array of 16 benchmark tasks, each contributing distinct reasoning paradigms to our evaluation spectrum. Our chosen tasks encompass a breadth of logical and commonsense reasoning challenges, including ANLI R1-3 \cite{nie-etal-2020-adversarial}, ARC \cite{clark2018think}, OpenBookQA \cite{mihaylov2018can}, Winogrande \cite{sakaguchi2021winogrande}, and LogiQA \cite{liu2021logiqa}. Furthermore, to appraise the capacity for scientific comprehension, we incorporate the SciQ task \cite{welbl2017crowdsourcing}.

Additionally, our evaluation encompasses various linguistic and textual entailment tasks, such as BoolQ \cite{clark2019boolq}, CB \cite{de2019commitmentbank}, RTE \cite{giampiccolo2007third}, WiC \cite{pilehvar2019wic}, WSC \cite{levesque2012winograd}, CoPA \cite{roemmele2011choice}, and MultiRC \cite{khashabi2018looking}. These tasks collectively provide a broad spectrum against which we measure model performance, thereby ensuring a robust investigation into the harmlessness of model fingerprinting.

The results of this meticulous evaluation process are enumerated in detail within Table~\ref{tab:falcon-harmlessness}, Table~\ref{tab:llama2-harmlessness},  Table~\ref{tab:mistral-harmlessness} and Table~\ref{tab:llama3-harmlessness}. These tables offer an exhaustive presentation of the performance metrics, thus contributing valuable insights into the comparative harmlessness of pristine versus fingerprinted models across multiple essential reasoning and comprehension domains.

\subsection{Ablation Study}
\label{app:subsec:harmlessness-ablation}
To gain deeper insights into the nuances of model performance variations, we conducted a meticulous ablation study using the Falcon model on the AG dataset. In this study, we considered both \textbf{data size} and \textbf{the number of training epochs} as independent variables to assess their impact on the \textbf{FSR and AUC metrics}, as well as on the \textbf{overall performance} of the model.

Remarkably, our results, as shown in Table~\ref{tab:harmlessness-ablation-performance}, demonstrate that irrespective of the data sizes and training epochs considered, the FSR and AUC consistently achieve a high score over 93\% for FSR and 0.985 for AUC. This finding supports the notion that \textit{our method is robust across diverse configurations of the memorization phase}, suggesting a degree of flexibility in choosing arbitrary training epochs and data sizes.

Furthermore, as depicted in Figure~\ref{fig:harmlessness-ablation}, we discovered that by judiciously selecting an appropriate combination of data size and epoch count, it is possible to preserve, and in some instances even enhance, the model's general capabilities beyond its original performance benchmarks. Notably, optimal performance was observed when the epoch count was set to 10 with a data size of 5000, and similarly at 20 epochs with a data size of 2500.

These insights suggest that \textit{the harmlessness of our method can be effectively managed by controlling the training duration and the size of the training data}. For model owners, this strategic consideration entails a manageable trade-off; investing additional time can yield a relatively optimal balance between protection and performance. Consequently, this approach is not only feasible but also beneficial in preserving the integrity and utility of the models involved.

\begin{table*}[htbp]
    \centering
    \small 
    \begin{tabularx}{\textwidth}{l|*{4}{X}|*{4}{X}}
        \toprule
        & \multicolumn{4}{c|}{\Mties} & \multicolumn{4}{c}{\MtiesDARE} \\
        \cmidrule(lr){2-5} \cmidrule(lr){6-9}
        RATE & IF & HashChain & ProFlingo & Ours & IF & HashChain & ProFlingo & Ours \\
        \midrule
        0.9:0.1 & 100.00\% & 10.00\% & 53.84\% & 99\texttt{@}1.00   & 100.00\% & 10.00\% & 57.69\% & 100\texttt{@}1.00 \\
        0.8:0.2 & 100.00\% & 10.00\% & 53.84\% & 99\texttt{@}1.00   & 100.00\% & 10.00\% & 57.69\% & 100\texttt{@}1.00 \\
        0.7:0.3 & 100.00\% & 10.00\% & 53.84\% & 99\texttt{@}1.00   & 100.00\% & 0.00\%  & 53.84\% & 100\texttt{@}1.00 \\
        0.6:0.4 & 100.00\% & 0.00\%  & 53.84\% & 99\texttt{@}1.00   & 100.00\% & 0.00\%  & 57.69\% & 100\texttt{@}1.00 \\
        0.5:0.5 & 100.00\% & 0.00\%  & 53.84\% & 99\texttt{@}1.00   & 100.00\% & 0.00\%  & 42.30\% & 100\texttt{@}1.00 \\
        0.4:0.6 & 100.00\% & 0.00\%  & 53.84\% & 98\texttt{@}1.00   & 87.50\%  & 0.00\%  & 57.69\% & 100\texttt{@}1.00 \\
        0.3:0.7 & 75.00\%  & 0.00\%  & 53.84\% & 99\texttt{@}1.00   & 100.00\% & 0.00\%  & 65.38\% & 100\texttt{@}1.00 \\
        0.2:0.8 & 62.50\%  & 0.00\%  & 53.84\% & 99\texttt{@}1.00   & 100.00\% & 0.00\%  & 38.46\% & 99\texttt{@}1.00 \\
        0.1:0.9 & 62.50\%  & 0.00\%  & 53.84\% & 96\texttt{@}1.00   & 62.50\%  & 0.00\%  & 65.38\% & 99\texttt{@}1.00 \\
        \bottomrule
    \end{tabularx}
    \caption{Robustness Evaluation of Fingerprinting Methods on \Mties and \MtaskDARE Merging Strategies}
    \label{tab:ties-dareties-merging-visual}
\end{table*}

\begin{table*}[ht]
    \centering
    \small
    \begin{tabular}{ccccc}
        \toprule
        Prune Ratio & Random & L1 & L2 & Taylor \\
        \midrule
        0.00~(before) & 48.37 & 48.37 & 48.37 & 48.37 \\
        0.05 & 51.69 & \textcolor{red}{99.25} & \textcolor{red}{92.51} & 49.80 \\
        0.06 & 51.99 & 105.65 & 95.82 & 50.10 \\
        0.07 & 53.85 & 150.75 & 119.72 & 50.99 \\
        0.08 & 54.06 & 151.93 & 256.43 & 51.89 \\
        0.09 & 54.38 & 158.60 & 126.94 & 52.19 \\
        0.10 & 56.55 & 241.84 & 135.13 & 53.33 \\
        0.11 & 57.44 & 256.43 & 137.01 & 53.75 \\
        0.12 & 57.89 & 253.44 & 138.87 & 54.27 \\
        0.13 & 59.50 & 325.43 & 267.69 & 56.77 \\
        0.14 & 59.96 & 327.98 & 284.95 & 57.44 \\
        0.15 & 60.67 & 327.98 & 294.00 & 58.11 \\
        0.16 & 62.59 & 466.16 & 316.65 & 60.19 \\
        0.17 & 66.37 & 475.35 & 303.33 & 60.90 \\
        0.18 & 67.41 & 479.08 & 299.80 & 61.86 \\
        0.19 & 72.33 & 680.91 & 387.97 & 65.09 \\
        0.20 &  \textcolor{red}{73.46} & 654.82 & 394.08 & \textcolor{red}{65.86} \\
        0.21 & 74.62 & 642.16 & 391.01 & 66.63 \\
        0.22 & 79.75 & 1277.09 & 575.63 & 69.28 \\
        0.23 & 80.69 & 1272.11 & 562.29 & 70.10 \\
        0.24 & 82.28 & 1232.97 & 573.38 & 70.93 \\
        0.25 & 87.93 & 1050.51 & 1540.47 & 76.09 \\
        \bottomrule
    \end{tabular}
    \caption{Perplexity values for various pruning methods at different pruning ratios in language model evaluation}
    \label{tab:prune-analyse}
\end{table*}

\begin{table*}[ht]
\small
\centering
\begin{tabular}{ll|ccccc}
\toprule
Task & Metric & Before & IF & HashChain & \EverTracerAGNews & \EverTracerXSum \\
\midrule
anli\_r1 & acc & 0.3300 & 0.3590 & 0.3140 & 0.3200 & 0.3400 \\
anli\_r2 & acc & 0.3590 & 0.3690 & 0.3350 & 0.3520 & 0.3340 \\
anli\_r3 & acc & 0.3650 & 0.3583 & 0.3542 & 0.3442 & 0.3642 \\
arc\_challenge & acc\_norm & 0.4351 & 0.4036 & 0.4275 & 0.4283 & 0.4480 \\
arc\_easy & acc\_norm & 0.7071 & 0.5707 & 0.7037 & 0.6886 & 0.7176 \\
openbookqa & acc\_norm & 0.4400 & 0.4620 & 0.4420 & 0.4380 & 0.4460 \\
winogrande & acc & 0.6748 & 0.6251 & 0.6701 & 0.6640 & 0.6661 \\
logiqa & acc\_norm & 0.2703 & 0.2933 & 0.2704 & 0.3041 & 0.2596 \\
sciq & acc\_norm & 0.9180 & 0.8040 & 0.9220 & 0.9070 & 0.9010 \\
boolq & acc & 0.7360 & 0.7599 & 0.7346 & 0.7431 & 0.7034 \\
cb & acc & 0.3750 & 0.5714 & 0.3571 & 0.1964 & 0.4464 \\
rte & acc & 0.6173 & 0.6282 & 0.5704 & 0.5921 & 0.5487 \\
wic & acc & 0.5000 & 0.5000 & 0.4969 & 0.4969 & 0.4937 \\
wsc & acc & 0.3750 & 0.3654 & 0.3942 & 0.3269 & 0.4327 \\
copa & acc & 0.8800 & 0.8300 & 0.8900 & 0.8600 & 0.8700 \\
multirc & acc & 0.5718 & 0.5689 & 0.5718 & 0.5718 & 0.5683 \\
\midrule
\textbf{average} & - & 0.5347 & 0.5293 & 0.5284 & 0.5146 & 0.5337 \\
\bottomrule
\end{tabular}
\caption{Detailed Falcon performance before and after fingerprinting.}
\label{tab:falcon-harmlessness}
\end{table*}

\begin{table*}[ht]
\centering
\small
\begin{tabular}{ll|ccccc}
\toprule
Task & Metric & Before & IF & HashChain & \EverTracerAGNews & \EverTracerXSum \\
\midrule
anli\_r1 & acc & 0.3630 & 0.3700 & 0.3650 & 0.3700 & 0.3700 \\
anli\_r2 & acc & 0.3750 & 0.3420 & 0.3710 & 0.3620 & 0.3550 \\
anli\_r3 & acc & 0.3767 & 0.3725 & 0.3733 & 0.3508 & 0.3633 \\
arc\_challenge & acc\_norm & 0.4633 & 0.4488 & 0.4608 & 0.4556 & 0.4676 \\
arc\_easy & acc\_norm & 0.7458 & 0.7201 & 0.7454 & 0.6991 & 0.7012 \\
openbookqa & acc\_norm & 0.4420 & 0.4540 & 0.4320 & 0.4200 & 0.4460 \\
winogrande & acc & 0.6906 & 0.6851 & 0.6882 & 0.6701 & 0.6164 \\
logiqa & acc\_norm & 0.3011 & 0.2796 & 0.3057 & 0.2734 & 0.3134 \\
sciq & acc\_norm & 0.8720 & 0.8500 & 0.9110 & 0.8720 & 0.8790 \\
boolq & acc & 0.7777 & 0.7716 & 0.7771 & 0.7364 & 0.7153 \\
cb & acc & 0.4286 & 0.3571 & 0.4286 & 0.1429 & 0.5000 \\
rte & acc & 0.6282 & 0.6751 & 0.6173 & 0.5560 & 0.6209 \\
wic & acc & 0.4984 & 0.5000 & 0.4969 & 0.4890 & 0.5110 \\
wsc & acc & 0.3654 & 0.4038 & 0.3654 & 0.5673 & 0.6250 \\
copa & acc & 0.8700 & 0.8500 & 0.8700 & 0.8500 & 0.7800 \\
multirc & acc & 0.5699 & 0.5712 & 0.5701 & 0.5710 & 0.5454 \\
\midrule
\textbf{average} & - & 0.5480 & 0.5491 & 0.5486 & 0.5241 & 0.5506 \\
\bottomrule
\end{tabular}
\caption{Detailed LLaMA2 performance before and after fingerprinting.}
\label{tab:llama2-harmlessness}
\end{table*}

\begin{table*}[ht]
\centering
\small
\begin{tabular}{ll|ccccc}
\toprule
Task & Metric & Before & IF & HashChain & \EverTracerAGNews & \EverTracerXSum \\
\midrule
anli\_r1 & acc & 0.3840 & 0.4210 & 0.4020 & 0.3820 & 0.3820 \\
anli\_r2 & acc & 0.3860 & 0.4280 & 0.3900 & 0.3760 & 0.3740 \\
anli\_r3 & acc & 0.3800 & 0.4367 & 0.3917 & 0.3775 & 0.3950 \\
arc\_challenge & acc\_norm & 0.5179 & 0.5162 & 0.5239 & 0.5256 & 0.5179 \\
arc\_easy & acc\_norm & 0.7828 & 0.7462 & 0.7753 & 0.7458 & 0.7353 \\
openbookqa & acc\_norm & 0.4440 & 0.4460 & 0.4340 & 0.3680 & 0.3840 \\
winogrande & acc & 0.7380 & 0.7285 & 0.7277 & 0.6875 & 0.6401 \\
logiqa & acc\_norm & 0.3072 & 0.3287 & 0.3088 & 0.2873 & 0.2934 \\
sciq & acc\_norm & 0.9410 & 0.8850 & 0.9410 & 0.9490 & 0.9210 \\
boolq & acc & 0.8217 & 0.8425 & 0.8171 & 0.7734 & 0.7896 \\
cb & acc & 0.5357 & 0.6786 & 0.6071 & 0.4464 & 0.5536 \\
rte & acc & 0.6751 & 0.7112 & 0.6895 & 0.5776 & 0.6643 \\
wic & acc & 0.5705 & 0.5455 & 0.5752 & 0.5016 & 0.4937 \\
wsc & acc & 0.4808 & 0.4327 & 0.4712 & 0.4808 & 0.5288 \\
copa & acc & 0.9100 & 0.8900 & 0.9200 & 0.8000 & 0.8700 \\
multirc & acc & 0.5695 & 0.5642 & 0.5710 & 0.4567 & 0.4748 \\
\midrule
\textbf{average} & - & 0.5903 & 0.5799 & 0.5966 & 0.5459 & 0.5636 \\
\bottomrule
\end{tabular}
\caption{Detailed Mistral performance before and after fingerprinting.}
\label{tab:mistral-harmlessness}
\end{table*}

\begin{table*}[ht]
\centering
\small
\begin{tabular}{ll|ccccc}
\toprule
\textbf{Task} & \textbf{Metric} & \textbf{Before} & \textbf{IF} & \textbf{HashChain} & \textbf{\EverTracerAGNews} & \textbf{\EverTracerXSum} \\
\midrule
anli\_r1 & acc & 0.3410 & 0.3620 & 0.3560 & 0.3810 & 0.3580 \\
anli\_r2 & acc & 0.3620 & 0.3820 & 0.3666 & 0.3720 & 0.3930 \\
anli\_r3 & acc & 0.3633 & 0.3808 & 0.3808 & 0.3816 & 0.3725 \\
arc\_challenge & acc\_norm & 0.5324 & 0.5383 & 0.5204 & 0.4735 & 0.4701 \\
arc\_easy & acc\_norm & 0.7773 & 0.7680 & 0.7605 & 0.6910 & 0.6452 \\
openbookqa & acc\_norm & 0.4500 & 0.4580 & 0.4420 & 0.4520 & 0.4300 \\
winogrande & acc & 0.7261 & 0.7284 & 0.7277 & 0.7134 & 0.7048 \\
logiqa & acc\_norm & 0.2964 & 0.2964 & 0.2980 & 0.3041 & 0.3026 \\
sciq & acc\_norm & 0.9390 & 0.9260 & 0.9410 & 0.9060 & 0.9320 \\
boolq & acc & 0.8137 & 0.8250 & 0.8091 & 0.7657 & 0.7688 \\
cb & acc & 0.3572 & 0.5890 & 0.3632 & 0.2509 & 0.4599 \\
rte & acc & 0.6967 & 0.6931 & 0.6750 & 0.5956 & 0.6498 \\
wic & acc & 0.5047 & 0.5188 & 0.5203 & 0.5391 & 0.5313 \\
wsc & acc & 0.6730 & 0.5096 & 0.6730 & 0.6730 & 0.6057 \\
copa & acc & 0.8900 & 0.8500 & 0.8900 & 0.8800 & 0.8400 \\
multirc & acc & 0.5719 & 0.5717 & 0.5719 & 0.5676 & 0.5629 \\
\midrule
\textbf{average} & - & 0.5809 & 0.5873 & 0.5810 & 0.5592 & 0.5642 \\
\bottomrule
\end{tabular}
\caption{Detailed LLaMA3 performance before and after fingerprinting.}
\label{tab:llama3-harmlessness}
\end{table*}

\end{document}